\documentclass[11pt]{article}
\usepackage{a4wide,rotating,amsmath,amssymb}

\begin{document}
\def\mrm{\mathrm}
\newcommand{\Zo}{\mbox{$\protect {\rm Z}^0$}}
\newcommand{\etal}{\mbox{$et$ $al.$}}
\newcommand{\PhysLett}  {Phys.~Lett. }
\newcommand{\PRL} {Phys.~Rev.\ Lett. }
\newcommand{\PhysRep}   {Phys.~Rep. }
\newcommand{\PhysRev}   {Phys.~Rev. }
\newcommand{\NPhys}  {Nucl.~Phys. }
\newcommand{\NIM} {Nucl.~Instr.\ Meth. }
\newcommand{\CPC} {Comp.~Phys.\ Comm. }
\newcommand{\ZPhys}  {Z.~Phys. }
\newcommand{\IEEENS} {IEEE Trans.\ Nucl.~Sci. }
\newcommand{\JHEP} {JHEP }
\begin{titlepage}
\begin{center}{\large   EUROPEAN ORGANIZATION FOR NUCLEAR RESEARCH
}\end{center}\bigskip
\begin{flushright}
       CERN-PH-EP/2009-013   \\  12 June 2009 \\ Modified 20 September 2009 \\
\end{flushright}
\bigskip\bigskip\bigskip\bigskip\bigskip
\begin{center}{\huge\bf\boldmath   
$\Sigma^-$-antihyperon correlations in \Zo~decay
and investigation of the baryon production mechanism
}\end{center}\bigskip\bigskip
\begin{center}{\LARGE The OPAL Collaboration
}\end{center}\bigskip\bigskip
\bigskip\begin{center}{\large  Abstract}\end{center}

Data collected around $\sqrt{s}=$91 GeV by the OPAL experiment 
at the LEP e$^+$e$^-$ collider are used to study the mechanism
of baryon formation.
As the signature, the fraction of $\Sigma^-$ hyperons whose baryon 
number is compensated by the production of a
$\overline{\Sigma^-}, \overline{\Lambda}$ or 
$\overline{\Xi^-}$ antihyperon is determined.
The method relies entirely on quantum number correlations of the
baryons, and not rapidity correlations, making it more model
independent than previous studies.
Within the context of the JETSET implementation of the
string hadronization model, the diquark baryon production model 
without the popcorn mechanism is
strongly disfavored with a significance of 3.8 standard 
deviations including systematic uncertainties.
It is shown that previous studies of the popcorn mechanism with
$\Lambda \overline{\Lambda}$ and $\mrm{p} \pi \overline{\mrm{p}}$ 
correlations are not conclusive, if parameter uncertainties are 
considered.

\bigskip\bigskip\bigskip\bigskip
\bigskip\bigskip
\begin{center}{\large
(Submitted to Eur. Phys. J. C.)
}\end{center}
\end{titlepage}
\begin{center}{\Large        The OPAL Collaboration
}\end{center}\bigskip
\begin{center}{
G.\thinspace Abbiendi$^{  2}$,
C.\thinspace Ainsley$^{  5}$,
P.F.\thinspace {\AA}kesson$^{  7}$,
G.\thinspace Alexander$^{ 21}$,
G.\thinspace Anagnostou$^{  1}$,
K.J.\thinspace Anderson$^{  8}$,
S.\thinspace Asai$^{ 22}$,
D.\thinspace Axen$^{ 26}$,
I.\thinspace Bailey$^{ 25}$,
E.\thinspace Barberio$^{  7,   p}$
% ...
}\end{center}\bigskip
\bigskip
$^{  1}$School of Physics and Astronomy, University of Birmingham,
Birmingham B15 2TT, UK
\newline
$^{  2}$Dipartimento di Fisica dell' Universit\`a di Bologna and INFN,
I-40126 Bologna, Italy
\newline
% ...
$^{  5}$Cavendish Laboratory, Cambridge CB3 0HE, UK
\newline
% ...
$^{  7}$CERN, European Organisation for Nuclear Research,
CH-1211 Geneva 23, Switzerland
\newline
$^{  8}$Enrico Fermi Institute and Department of Physics,
University of Chicago, Chicago IL 60637, USA
\newline
% ...
$^{ 21}$Department of Physics and Astronomy, Tel Aviv University,
Tel Aviv 69978, Israel
\newline
$^{ 22}$International Centre for Elementary Particle Physics and
Department of Physics, University of Tokyo, Tokyo 113-0033, and
Kobe University, Kobe 657-8501, Japan
\newline
% ...
$^{ 25}$University of Victoria, Department of Physics, P O Box 3055,
Victoria BC V8W 3P6, Canada
\newline
$^{ 26}$University of British Columbia, Department of Physics,
Vancouver BC V6T 1Z1, Canada
\newline
% ...
$^{  p}$ now at The University of Melbourne, Victoria, Australia
\newline
% ...
$^{  s}$ now at University of Antwerpen, Physics Department,B-2610 Antwerpen, 
Belgium; supported by Interuniversity Attraction Poles Programme -- Belgian
Science Policy
\newline
% ...
\section{Introduction}
\label{sect:intro}

The formation of baryons within a jet of hadrons has proved difficult 
to model and is still not well understood. While the shape of momentum
spectra can be derived from QCD with the modified leading logarithmic
approximation together with parton-hadron duality~\cite{lnxp},
more complex observables like correlations have not been derived
from first principles.

\begin{figure}
\centering
\includegraphics[width=0.9\textwidth]{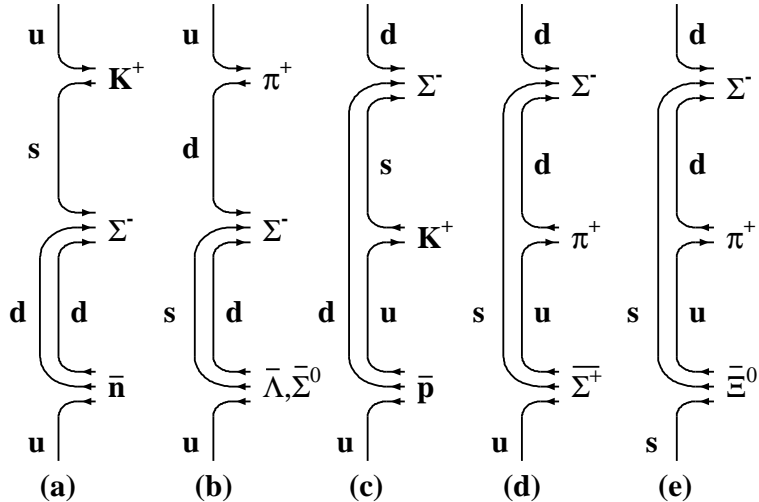}
\caption{ \sl Baryon production in the diquark model without
(a,b) and with (c-e) the popcorn effect.}
\label{fig:quarkflow}
\end{figure}

Several physical models like the thermodynamic~\cite{thermodyn},
cluster fragmentation~\cite{herwig}
or string fragmentation~\cite{stringmodel,diquark} models, have been 
developed to describe baryon production.
Of these, the most successful is string 
fragmentation, based on the creation of diquark-antidiquark pairs from 
the vacuum as illustrated in figure~\ref{fig:quarkflow}.
In a production chain of particles along the string, a baryon and an 
antibaryon can be produced in immediate succession 
(figs.~\ref{fig:quarkflow}a,b) or else one or more intermediate 
mesons can be produced between them (figs.~\ref{fig:quarkflow}c-e).
This production of intermediate mesons, referred to as the
popcorn effect~\cite{popcorn,mops}, is included as an option in the 
Monte Carlo event generators JETSET and PYTHIA ~\cite{jetset,pythia61} 
and can be steered with a free parameter.

Past experimental investigations of the popcorn effect made use of 
rapidity ordering of hadrons in the fragmentation chain.
Intermediate mesons modify the rapidity difference 
between associated baryons and antibaryons. Rapidity correlations 
between $\Lambda \; \overline{\Lambda}$ pairs produced in \Zo~decays have 
been studied by several LEP collaborations and the conclusion was 
that best agreement 
between the experiments and the JETSET Monte Carlo model ~\cite{jetset} 
was obtained with the popcorn effect 
included~\cite{lambda-pair-aleph}-\cite{lambda-pair-opal}.

A contradictory result was reported by the DELPHI collaboration
~\cite{ppip-delphi}. Their measurement is based on triple correlations between
a proton, an antiproton and a charged pion close in rapidity.
Because the popcorn effect enhances the pion density
in the rapidity interval between the proton and antiproton,
the minimum rapidity difference between a pion and a proton
was compared for the particle orderings
$\mrm{p} \pi \overline{\mrm{p}}$ and $\pi \mrm{p} \overline{\mrm{p}}$.
The measurement indicated that the rapidity rank correlations could be
reproduced without the popcorn effect. The contribution of events
with popcorn produced mesons was reported to be less than 
15\% at 90\% confidence level.
Insufficient modeling of the fragmentation dynamics could
not, however, be excluded~\cite{ppip-delphi}.

In this analysis, the popcorn mechanism is investigated
in a different way, 
by tagging rare baryons and measuring the quantum numbers 
of correlated antibaryons. By not relying on rapidity differences,
we obtain results that are more model-independent than previous 
studies.
The data were collected with the OPAL experiment at the 
LEP e$^+$e$^-$ collider at CERN. 
Especially suitable for our purposes is the $\Sigma^-$ hyperon. If
a $\Sigma^-$ is produced, its baryon number and strangeness are 
compensated either by an antinucleon and a kaon, as 
illustrated in figure~\ref{fig:quarkflow}a, or by an associated 
antihyperon. 
The case of $\overline{\Lambda}$ or $\overline{\Sigma^0}$ production 
without the popcorn effect is shown in figure~\ref{fig:quarkflow}b. 
The same graph, with the bottom-most u-quark replaced 
by a d- or s-quark, describes associated $\overline{\Sigma^-}$ and 
$\overline{\Xi^-}$ production.
Associated production of a $\overline{\Sigma^+}$ 
and $\overline{\Xi^0}$ antihyperon with a $\Sigma^-$
can only occur through the popcorn mechanism 
(figs.~\ref{fig:quarkflow}d,e).
This makes $\Sigma^- \overline{\Sigma^+}$ and $\Sigma^- \overline{\Xi^0}$
correlations ideal tools to study the popcorn effect.
Unfortunately, the production rate of $\Sigma^-$ hyperons and the 
probabilities for simultaneous reconstruction of 
$\Sigma^-$ and $\overline{\Sigma^+}$ or $\Sigma^-$ and 
$\overline{\Xi^0}$ particles are too small to make such an analysis 
feasible, given the available data statistics.
The equivalent analysis cannot be performed with tagged $\Lambda$ 
hyperons because in this case $\overline{\Sigma^+}$ and 
$\overline{\Xi^0}$ antihyperons can be produced without the popcorn 
effect. Alternatively one can measure the fraction 
\begin{equation}
\label{eq:Fdef}
F_{\overline{H}} =
F_{\Sigma^-,\overline{\Sigma^-}} +
F_{\Sigma^-,\overline{\Lambda}} +
F_{\Sigma^-,\overline{\Xi^-}} 
\end{equation}
of $\Sigma^-$ hyperons accompanied by a
$\overline{\Sigma^-}$ $(F_{\Sigma^-,\overline{\Sigma^-}}$), 
a $\overline{\Lambda}$ ($F_{\Sigma^-,\overline{\Lambda}}$) or 
a $\overline{\Xi^-}$ ($F_{\Sigma^-,\overline{\Xi^-}}$).
These correlations can occur in the popcorn model but are more
likely in the diquark model (fig.~\ref{fig:quarkflow}b) and their 
rate is thus a sensitive measure of the baryon production 
mechanism.

The exact definition of the three correlations 
$F_{\Sigma^-,\overline{k}}$ in the sum (\ref{eq:Fdef})
needs to account for the possibility that an event may contain more
than one $\Sigma^-$ hyperon, other additional hyperons, 
or more than one antihyperon.
More generally, one can consider an arbitrary particle $k$ and its
antiparticle $\overline{k}$.
If the number of  $\Sigma^-$ hyperons in an event is larger than 1,
all combinations of \mbox{$\Sigma^- \; \overline{k}$} and 
$\Sigma^- \; k$ pairs are counted. 
Denoting the total rates of $\Sigma^-$-antiparticle and 
$\Sigma^-$-particle pairs by $R_{\Sigma^-,\overline{k}}$ 
and $R_{\Sigma^-,k}$, respectively, and 
the total $\Sigma^-$ rate by $R_{\Sigma^-}$, 
the fraction $F_{\Sigma^-,\overline{k}}$ can be written as
\begin{equation}
\label{eq:pair-fraction}
F_{\Sigma^-,\overline{k}} = \frac 
{R_{\Sigma^-,\overline{k}} - R_{\Sigma^-,k}}
{R_{\Sigma^-}} \ .
\end{equation} 
This definition implies that the $n_{\overline{k}}$ antiparticles 
$\overline{k}$ in an event enter $n_{\overline{k}}$\,$n_{\Sigma^-}$ 
times. The $\Sigma^- \; \Sigma^-$ pairs are counted 
$n_{\Sigma^-}$\,$(n_{\Sigma^-}-1)$ times. 
In the data analysis, this multiple counting  
is not an issue because the number of reconstructed $\Sigma^-$ 
hyperons per event is very small.
Baryon number conservation ensures that 
$\sum_{\overline{k}} F_{\Sigma^-,\overline{k}}=1$, if the sum
extends over all antibaryons, including the antihyperons 
and antinucleons.
It is the understanding throughout this paper
that the charge conjugated channels are included.
Thus, the rates $R_{\Sigma^-,k}$ contain all like-sign 
pairs $\Sigma^- \;k$ and $\overline{\Sigma^-} \; \overline{k}$,
the rates $R_{\Sigma^-, \overline{k}}$ contain all unlike-sign pairs
$\Sigma^- \; \overline{k}$ and $\overline{\Sigma^-} \; k$,
and $\overline{\Sigma^-}$ antihyperons are included in 
$R_{\Sigma^-}$. 

It is evident from figure~\ref{fig:quarkflow}c
that kaons created through the popcorn effect reduce $F_{\overline{H}}$.
The  PYTHIA Monte Carlo program without the popcorn effect,
tuned to reproduce the observed baryon rates and momentum spectra, 
predicts $F_{\overline{H}}\approx 0.9$, as will be shown later.
The other extreme is a model in which baryon number and strangeness
are compensated statistically, i.e. from conservation laws alone.
The ratio of weakly decaying hyperon to total baryon production in 
\Zo~decays is approximately 0.22~\cite{PDGmanual}.
Because there must be at least one antibaryon in the rest of the event,
if the $\Sigma^-$ is detected, and the tagging biases the 
number of strange valence antiquarks, the lower bound will be a bit 
larger: $F_{\overline{H}} > 0.22$.
A more rigorous calculation can be performed with the thermodynamic 
model for particle production. 
The advanced version of the model, based on the
microcanonical ensemble, yields
$F_{\overline{H}}=0.23$ for an initial system 
without strangeness~\cite{thermodyn2}, close to the simple
estimate of the lower bound.
 
\section{Event selection}
\subsection{Event topologies}
\label{sect:selection}

In any experiment with a sufficiently large tracking device
$\Sigma^-$ hyperons can be identified by track kinks
from $\Sigma^- \rightarrow \mrm{n} \pi^-$ decays.
At LEP energies, the efficiency is much less 
than 100\%, because the decay vertex lies often outside the fiducial 
volume for its reconstruction. In principle, correlated decays 
$\overline{\Sigma^-} \rightarrow \overline{\mrm{n}} \pi^+$  and 
$\overline{\Xi^-} \rightarrow \overline{\Lambda} \pi^+$ could 
be reconstructed using track kinks, too,
but the efficiency for the exclusive reconstruction 
of $\Sigma^- \; \overline{\Sigma^-}$ or $\Sigma^- \; \overline{\Xi^-}$
pairs is small.

In this work, only the $\Sigma^-$ hyperons were reconstructed
exclusively. Two signatures for correlated antihyperons were used:
\begin{enumerate}
\item
$\overline{\Lambda}$ hyperons were reconstructed by 
analyzing their so-called V$^0$ topology from the decay
$\overline{\Lambda} \rightarrow \overline{\mrm{p}} \pi^+$.
The decay vertex and the flight direction of the 
$\overline{\Lambda}$ candidates allow 
the $\overline{\Lambda}$ impact parameters $d_0$ with 
respect to the beam line to be computed.
Direct $\overline{\Lambda}$ production, 
including the contribution of decays from $\overline{\Sigma^0}$ hyperons, 
is characterized by low impact parameters $d_0$, while
large impact parameters indicate a preceding weak decay and are a 
signature for $\overline{\Xi}$ decays. 
\item
Charged pions with significant impact parameters
are a signature for weak decays of arbitrary antihyperons.
An inclusive sample of tracks with large $d_0$ values,
consistent with a pion interpretation, was 
selected. Throughout this paper, this data set is referred 
to as the sample of displaced tracks. This
sample has a large background, and different
antihyperon species contribute with different weights, because 
the number of decay pions per antihyperon is 1 for
$\overline{\Sigma^-}$, 0.64 for $\overline{\Lambda}$ and
1.64 for chain decays of $\overline{\Xi^-}$.
\end{enumerate}

The correlated $\Sigma^-\overline{\Lambda}$ candidate sample gives 
the numbers of true $\Sigma^- \; \overline{\Lambda}$ and
$\Sigma^- \; \overline{\Xi^-}$  pairs. The fraction 
$F_{\overline{\Sigma^-}}$ can be extracted from the displaced track
sample by a weighted subtraction of the
$\Sigma^- \; \overline{\Lambda}$ and
$\Sigma^- \; \overline{\Xi^-}$ contributions.  

\subsection{Experiment and data sets}

All data taken by the OPAL experiment~\cite{detector}
in the \Zo~energy region during the years 1991 to 2000 were analyzed 
to measure the correlation. The OPAL experiment had
nearly complete solid angle coverage and excellent hermeticity.
The innermost part of the central tracking detector was a high-resolution
silicon microvertex detector, which immediately surrounded the 
beam-pipe~\cite{strip}. It was followed by a high-precision 
vertex drift chamber, a large-volume jet chamber~\cite{jetchamber}, 
and $z$-chambers, all in a uniform 0.435 T axial magnetic field. 
In this work, the outer detector parts as well as the forward detector 
system were needed for triggering and identification of multihadronic
events only. The criteria for multihadronic event selection have 
been described elsewhere~\cite{MH-selection}. 

The present analysis is entirely based on the central tracking
system~\cite{detector,jetchamber}. 
For candidates to be accepted, all central wire chambers and the
microvertex detector were required to be fully operational.
The data sample for this analysis consists of 3.97 million events.

The identification of strange particles is based on earlier 
work~\cite{sigma-OPAL,strange-baryons-opal}.
The innermost sense wires of the jet chamber had a distance of 25.5 cm 
from the beam spot, the wire-to-wire distance was 1 cm and there were
159 sensitive radial layers. The requirement of the pattern recognition 
program was the existence of at least 12 hits in the jet chamber. 
This makes it possible to identify $\Sigma^-$ 
hyperons with decay lengths larger than 36 cm.

The $z$ coordinate, along an axis parallel to the electron beam, 
was measured with a precision of 700 $\mu$m with
the stereo wires of the vertex chamber and 100 to 350 $\mu$m with the
$z$ chambers. For tracks leaving the drift chamber at the side cones, 
the $z$ coordinates of the exit points can be computed from the radius
of the last wire with a hit. At the beam spot a constraint can be set 
using the bunch length of the beam.
Inside the jet chamber, the $z$ coordinates were measured with the
charge division method with a resolution of 6 cm. This is
one limiting factor for the kinematical reconstruction of $\Sigma^-$ 
hyperons to be discussed later. 

The quality of the impact parameter measurement is directly connected to 
the detector resolutions in the $(r,\varphi)$ plane perpendicular 
to the beam axis. These resolutions were
5 $\mu$m to 10 $\mu$m for the microvertex detector~\cite{strip}, 
55 $\mu$m for the vertex chamber and, on average, 135 $\mu$m for the 
jet chamber~\cite{detector}.   

To study the detector response, the selection 
was applied also to Monte Carlo samples used before at OPAL.
These were generated with the JETSET7.3 and JETSET 7.4 programs, 
followed by a full detector simulation ~\cite{gopal}.
The steering parameters for the generator are
given in \cite{opaltunejt73,opaltunejt74}.
The subset of parameters relevant for this analysis is discussed in 
section~\ref{sect:modelpars} and the numerical values are given in
table~\ref{table:MCparameter} (appendix).
In total, the Monte Carlo samples consisted of 4.65 million
multihadronic \Zo~decays.
 
\subsection{$\Sigma^-$ Selection}
\begin{table}[t]
\begin{center}
\begin{tabular}{|r|c|}
\hline
variable                  &  condition                      \\
\hline
\hline
transverse momenta of the decaying & \\
and the daughter particle & \raisebox{1.5ex}[0cm][0cm]
{$p_{T1} > 0.15$ GeV/c, $p_{T2} > 0.10$ GeV/c} \\
\hline
total momentum of the decaying particle & $p_1 > 2$ GeV/c   \\
\hline
impact parameter of the primary track    & $d_0 < 5$ cm \\
\hline
endpoint radius of the primary track     & $r_{E1} < 170$ cm \\
\hline
maximum gap between the primary and     & \\ 
the secondary track in the $(r,\varphi)$ plane & 
\raisebox{1.5ex}[0cm][0cm]{$d_{12}< 10$ cm} \\
\hline
number of secondary tracks     & 1     \\
\hline
particle charges               & equal \\
\hline
arc distance between the intersection point & \\
and the end points of tracks 1,2 in $(r,\varphi)$ & 
\raisebox{1.5ex}[0cm][0cm]{$\mu_1 > -7$ cm, $\mu_2 < +7$ cm} \\
\hline
radius of the intersection point   & $r_{Vtx} > 35$ cm \\
\hline
distance of the intersection point &  \\
from the end plate in $z$ direction & 
\raisebox{1.5ex}[0cm][0cm]{$\Delta z_{Vtx} > 40$ cm} \\
\hline
$\chi^2$ for agreement of the two tracks in the & \\
$(r,\varphi)$ plane, using $(3 \times 3)$ error matrix &
\raisebox{1.5ex}[0cm][0cm]{$\chi^2> 2000$} \\
\hline
ratio of the track length to the & \\
momentum of the decaying particle & 
\raisebox{1.5ex}[0cm][0cm]{$l_1/p_1 < 15$ cm/(GeV/c)} \\
\hline
\end{tabular}
\end{center}
\caption{ \sl Selection of track pairs forming a decay vertex
(for a more detailed description of the variables, 
see~\cite{sigma-OPAL}). }
\label{table:kink-selection}
\end{table}

The production cross section of $\Sigma^-$ hyperons has previously
been measured by OPAL~\cite{sigma-OPAL}. Here the selection criteria to find
track pairs forming a decay vertex were slightly changed relative
to~\cite{sigma-OPAL} to improve the sensitivity of the analysis 
to correlations.
The applied cuts are summarized in table~\ref{table:kink-selection}.
They accommodate pattern recognition tolerances
and define a fiducial jet chamber volume to guarantee minimal hit
numbers for both tracks and to remove background from the end
plates. The ratio $l_1/p_1$ in the last line of 
table~\ref{table:kink-selection} is proportional to the decay 
time in the rest frame of the decaying particle. High values are
rejected to reduce the substantial background from kaon decays.

\newpage
Monte Carlo studies show that the following processes 
have to be considered as sources for $\Sigma^-$ candidates, others
being negligible~\cite{sigma-OPAL}: 
\begin{enumerate}
\item $\Sigma^- \rightarrow \mrm{n} \pi^-$; $\quad$ 
\item $\overline{\Sigma^+} \rightarrow \overline{\mrm{n}} \pi^-$; $\quad$
\item $\Xi^- \rightarrow \Lambda \pi^-$;
\item $\mrm{K^-} \rightarrow \pi^-$+neutrals; 
   $\mrm{K^- } \rightarrow \mu^-$+neutrals  
or $\mrm{K^- } \rightarrow e^-$+neutrals;
\item secondary reactions in the detector material and 
fake background. The last sample consists mainly of scattered particles.
\end{enumerate}
To determine the relative contributions of these processes 
to the observed data sample,
an unfolding procedure very similar to that used in~\cite{sigma-OPAL} 
was used.
The invariant mass $m_{\Sigma^-}$ and the decay angle 
$\theta^*$ of the pion candidate in the rest frame of the 
hypothetical $\Sigma^-$ particle were computed,
assuming that the unseen neutral particle
is a neutron. The angle $\theta^*$ is defined with respect to
the flight direction of the $\Sigma^-$ candidate at the decay vertex.

\begin{figure}[htb] 
\centering
\includegraphics[width=0.9\textwidth]{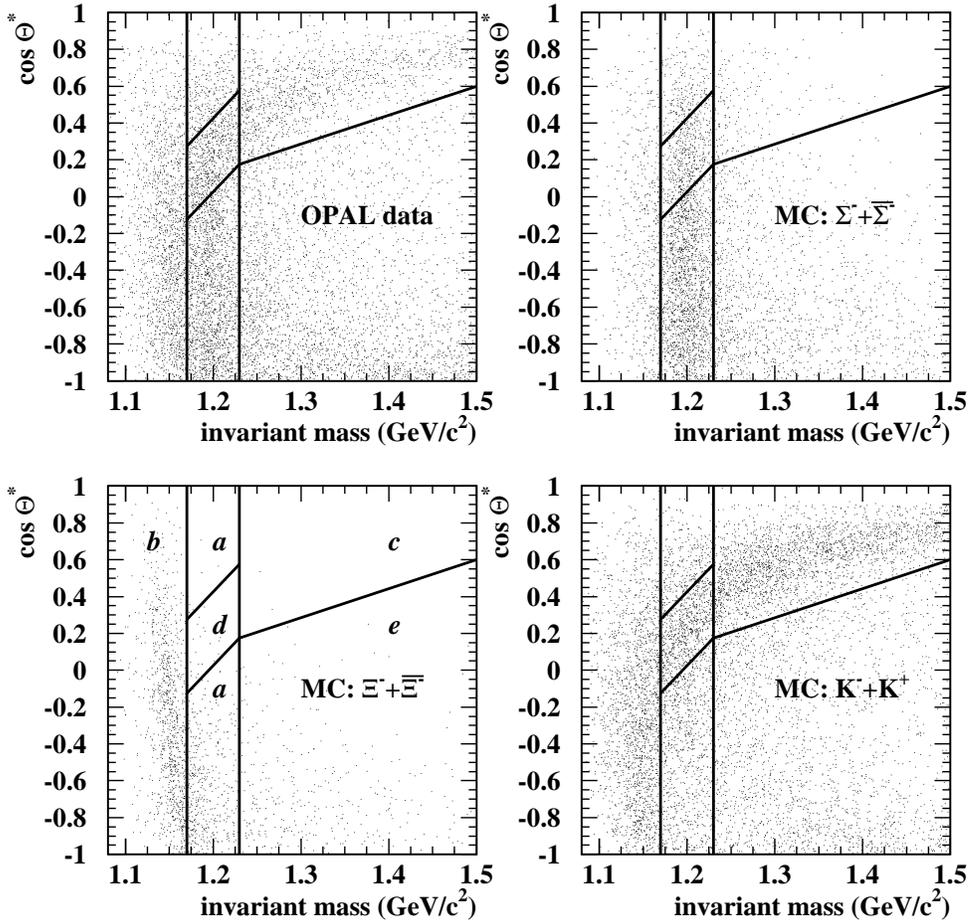}
\caption{\sl Definition of two-dimensional bins to disentangle
$\Sigma^-$ hyperons from background. The reconstructed mass is 
plotted versus the pion emission angle in the hypothetical 
center-of-mass system. Top left: data. Other plots:
Monte Carlo simulations for three particle classes as indicated.
The labeling of the two-dimensional bins is shown in the bottom 
left plot. The preferential bins for the kink sources are:
{\it a+d} for $\Sigma^-$ and $\overline{\Sigma^+}$, {\it b} for 
$\Xi^-$, {\it b,c} and {\it d}
for K$^-$ and {\it e} for background.}
\label{fig:mass-vs-cost}
\end{figure}

Figure \ref{fig:mass-vs-cost} shows scatter plots of these variables
for the data and three Monte Carlo sources of events with kinked tracks,
namely events with $\Sigma^-$, $\Xi^-$ or K$^-$ in the final state.
The background 5, not shown in 
figure~\ref{fig:mass-vs-cost}, is a smooth function 
of both observables and is largest in the backward direction
$\cos \theta^*=-1$. 

Five two-dimensional regions, denoted by $a,...e$, were introduced in the 
$(m_{\Sigma^-},\cos \theta^*)$ plane to enrich dedicated kink 
sources (see figure 2).
The bin number of a $\Sigma^-$ candidate was used as an observable 
in the unfolding procedure to be described in 
section~\ref{sect:analysis}.

In total, 16790 candidates were found in the two-dimensional 
plane in the mass range from threshold to 1.5 GeV. In the Monte Carlo 
sample 18754 kinks were identified.

\clearpage


\subsection{Selection of pions from weak decays}
\label{sect:pionselection}

Two pre-cuts were applied to select tracks with a good 
reconstruction quality: the transverse momentum  with respect to the 
beam had to be larger than 0.15 GeV/c and the track angle at the 
beam spot relative to the beam direction 
was restricted to the region $|\cos \theta| \le 0.80$.

To remove charged particles from charm or bottom decays, the
impact parameter $d_{0,\pi}$ with respect to the primary vertex
was required to be larger than 0.2 cm. 
This cut is the essential condition to define the sample of
displaced tracks. A lower value would be sufficient but 
does not improve the accuracy of the correlation analysis.

Pions were enriched using the specific energy loss measurement of
the central drift chamber~\cite{dedx}. 
A weight $w_{dE/dx}(\pi)$ was defined as the probability that the 
energy loss $dE/dx$ of a pion deviates from the median value
$<dE/dx>(\pi)$ by more than the measured difference from the median 
value. The applied condition was $w_{dE/dx}(\pi)>0.02$ and the 
number of hits contributing to this measurement had to be at least 20.

Finally, the angle between the momenta of
the pion candidate and the $\Sigma^-$ candidate at the primary vertex 
was required to be less than 90 degrees.
This hemisphere cut is motivated by the fact that it rarely
happens that baryon number and strangeness are compensated by
an antihyperon in the opposite event hemisphere. The cut
reduces the combinatorial background by a factor 2.

In total, 9965 correlated like-sign $\Sigma^- \pi^-$
and 11951 unlike-sign $\Sigma^- \; \pi^+$ pair candidates were 
selected with these cuts.
The corresponding results for the Monte Carlo sample are 10769
and 13818, respectively. If the total number of Monte Carlo 
track kinks is scaled to the observation,
the number of like-sign pairs in the data is well reproduced by the 
Monte Carlo, the difference being $(+3.4 \pm 1.4)$\%.
The observed unlike-sign minus like-sign difference, however, 
deviates from the prediction by $(-28 \pm 7)$\%.
This deficit, already visible at raw data level, indicates 
that the Monte Carlo sample contains too many correlated 
antihyperons and is the basis for the final result of this paper. 


\subsection{$\Lambda$ Selection}
\label{sect:lambdaselection}

The selection cuts to find $\Lambda$ decays in the
central drift chamber have been described
in~\cite{strange-baryons-opal}. The preselection cuts were relaxed
from those of~\cite{strange-baryons-opal}. In this analysis
all candidates with reconstructed masses between the $\mrm{p} \pi$ threshold 
and 1.20 GeV/c$^2$ were accepted. This larger mass window was needed
to study the non-$\Lambda$ background.

Two criteria were added. If the reconstructed $\Lambda$ flight
path points back to the $(r,\varphi)$ position of the hypothetical
$\Sigma$ decay kink within 2 degrees, it was assumed that the kink 
originated from a $\Xi^- \rightarrow \Lambda \pi^-$ decay 
and the $\Lambda$ candidate was dropped. The cut removed 
approximately 3/4 of the like sign 
$\Xi^- \; \Lambda \; (\overline{\Xi^-} \; \overline{ \Lambda})$-pairs
and reduced the self-correlation of $\Xi^-$ particles with their own 
decay $\Lambda$'s accordingly.

Secondly, the hemisphere cut applied to the decay pions was also applied
to the $\Lambda$'s. The angle between the flight directions of
$\Sigma$ and $\overline{\Lambda}$ at the primary vertex was required 
to be less than 90 degrees.

For the correlation analysis, the mass window was reduced to a 
$\pm 10$ MeV wide interval around the true $\Lambda$ mass.
Totals of 276 $\Sigma^- \Lambda$ \, 
$(\overline{\Sigma^-}\, \overline{\Lambda})$ and 
604 $\Sigma^- \; \overline{\Lambda}$ $(\overline{\Sigma^-} \; \Lambda)$
pair candidates passed all selection cuts. In the Monte Carlo sample
284 like-sign and 721 unlike-sign pairs were found.   
The observed unlike-sign minus like-sign difference is smaller than 
the Monte Carlo prediction by $(16 \pm 11)$\%, if normalized to the
number of observed kinks.

\section{\boldmath Correlation analysis} \label{sect:analysis}

\subsection{Unfolding of kink sources}

From the Monte Carlo sample one gets, for every kink source $i$
and every two-dimensional $(m_{\Sigma^-},\cos \theta^*)$ bin $j$, 
the number of accepted events $K^{(MC)}_i(j)$. 
The populations of the bins $j=${\it a,...e} in figure 2 are sensitive 
to the invariant mass and $\theta^*$ resolutions, which in turn depend
on the $z$ resolution of the central drift chamber.
The $z$ coordinate, however, is not well modeled in the Monte Carlo
simulation. Because the track end points in the jet chamber are well
known by other measurements, this mismodeling at the decay
vertex gives the dominant 
contribution to the systematic error of the 
$\Sigma^-$ rate~\cite{sigma-OPAL}.   
To correct the Monte Carlo program for it, the $z$ components 
of all $\Sigma$ momenta were modified according to 
$p_{z^,new}=p_{z,rec}+ c \cdot (p_{z,rec}-p_{z,true})$, where 
$p_{z,true}$ is the true momentum from the MC generator and
$p_{z,rec}$ the reconstructed momentum. The constant $c$ is one
common factor to be determined in the analysis.

The total number of kinks $K(j)$ expected in bin $j$ is given by
\begin{eqnarray}
\label{eq:kink-bin}
K(j) = \sum_i \xi_i \frac{N_{data}}{N_{MC}} \cdot K^{(MC)}_i(j) \ ,  
\end{eqnarray}
where $N_{data}$ and $N_{MC}$ are the total
number of multihadronic data and Monte Carlo events, respectively.
Incorrect Monte Carlo rates are corrected for by 
the five scaling factors $\xi_i$. 
If they are known, the true production rates per multihadronic event, 
$R_i$, can be computed for all sources $i$, for example
\begin{eqnarray}
\label{eq:sigma-rate}
R_{\Sigma^-} = \xi_{\Sigma^-} \cdot R_{\Sigma^-}^{(MC)} \quad
{\rm{with}} \quad R_{\Sigma^-}^{(MC)}= \sum_j K^{(MC)}_{\Sigma^-}(j)
\ . 
\end{eqnarray}

Contrary to~\cite{sigma-OPAL} the factors $\xi_i$ were treated
as momentum independent, because the analy\-sis of~\cite{sigma-OPAL} 
had shown that the modeling of the spectral shape was satisfactory.
The contribution from $\overline{\Sigma^+}$ hyperons to $K(j)$ is 
less than 20\% of that of the $\Sigma^-$ particles~\cite{sigma-OPAL}.
The ratio of the genuine production rates, approximately 1 due to
isospin symmetry, was fixed to the Monte Carlo prediction, 
so that $\xi_{\overline{\Sigma^+}}=\xi_{\Sigma^-}$. 
Four scaling factors $\xi_i$ were thus left for adjustment.
They were computed with a $\chi^2$-fit of the measured kink rates
to equation~({\ref{eq:kink-bin}}) for a given value of $c$.

\begin{figure}[ht] 
\centering
\includegraphics[width=0.9\textwidth]{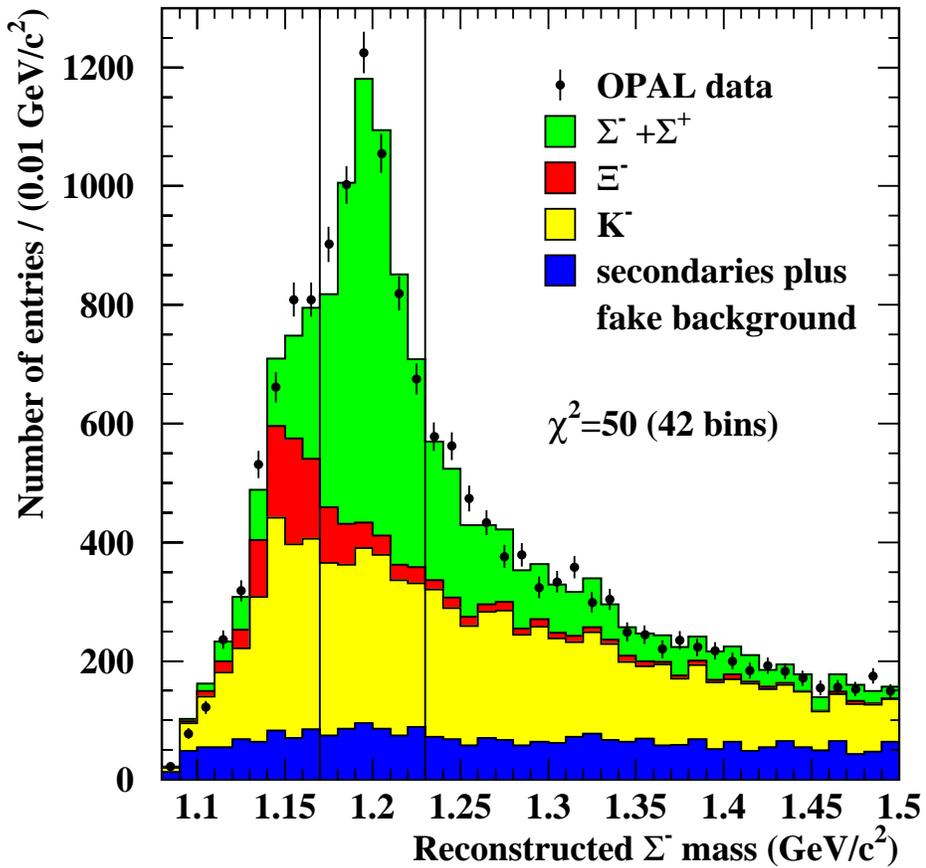}
\caption{ \sl Measured $\Sigma^-$ mass spectrum. Points: data.
Histograms: results of the fit, ordered according to the sources for
track kinks. The plotted errors are the statistical errors of the data.}
\label{fig:mass-sigma}
\end{figure}

\begin{figure}[htb] 
\centering
\includegraphics[width=0.9\textwidth]{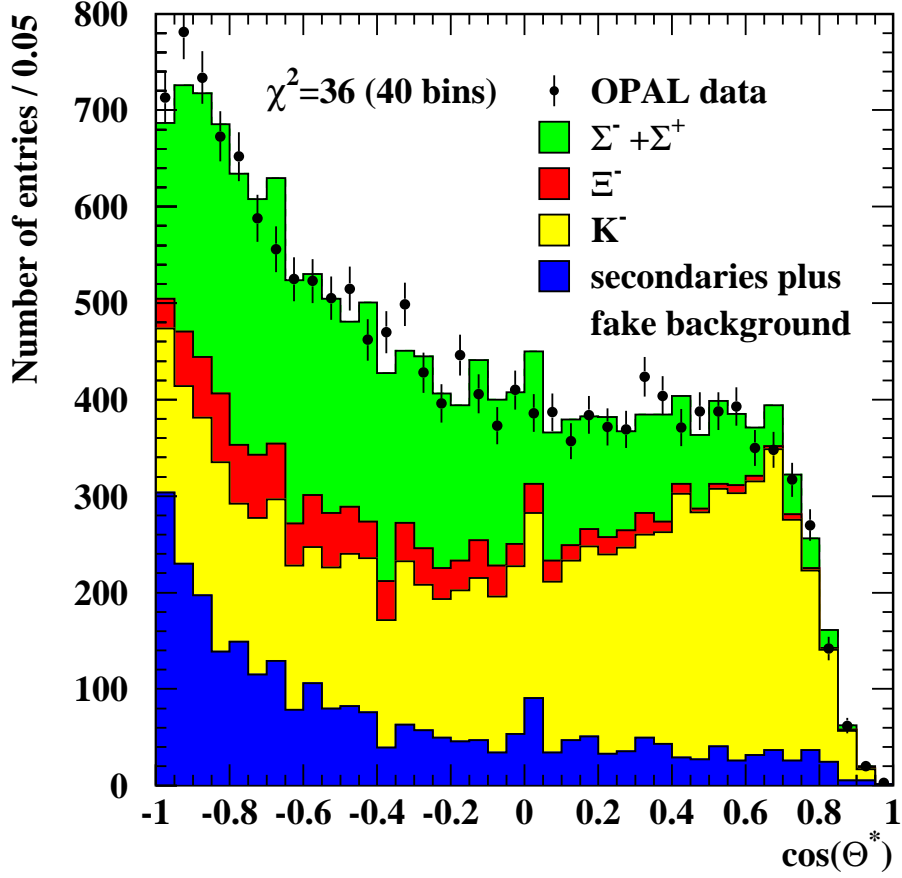}
\caption{ \sl  Cosine of the center-of-mass pion emission angle.
Points: data. Histograms: results of the fit, ordered according 
to the sources for track kinks. The plotted errors are the 
statistical errors of the data.}
\label{fig:costhstar-sigma}
\end{figure}

It was then checked whether the Monte Carlo simulation reproduces 
the reconstructed mass and $\cos \theta^*$ distributions.
Differing from our previous analysis~\cite{sigma-OPAL},   
the proportionality factor $c$ was chosen to get the lowest
$\chi^2$-sum for both distributions. The best overall agreement 
was found for $c=1.35$ with a one-sigma interval ranging from
1.32 to 1.45. The final fit result is shown in 
figs. \ref{fig:mass-sigma} and \ref{fig:costhstar-sigma}. 
The agreement is excellent. 
The error of $c$ is not included in the statistical errors of the
$\xi_i$. It is treated separately as a contribution to the systematic 
error.
\clearpage
\newpage
The $\Sigma^-+\overline{\Sigma^-}$ 
production rate was found to be $R_{\Sigma^-}=0.073 \pm 0.004$. 
According to~\cite{sigma-OPAL} a systematic error of 
$\pm 0.009$ has to be added. Within the total error, this result
is consistent with our published value $R_{\Sigma^-}=0.083 \pm 0.011$
and also with the world average 
$R_{\Sigma^-}=0.082 \pm 0.007$~\cite{PDGmanual}.
The difference is mainly due to the modified treatment of the 
$z$-resolution.  
\clearpage
\subsection{Correlated particle sources}

The correlated rates of kink-V$^0$ or kink-displaced-track
pairs were measured as a function of the bin number $j$ and 
the impact parameter $d_0$ of the correlated particle
relative to the beam line. The variable $d_0$ 
contains information on the lifetime of the parent particle
of the correlated particle in case of a preceding
weak decay, and is needed to disentangle correlated $\overline{\Lambda}$'s 
from $\overline{\Xi}$'s decaying into $\overline{\Lambda}$'s.

In the Monte Carlo simulation, nine sources have to be considered 
for the displaced tracks and correlated V$^0$'s:
\begin{enumerate}
\item $\overline{\Sigma^-} \rightarrow \overline{\mrm{n}} \pi^+$;
\item $\overline{\Sigma^+} \rightarrow \overline{\mrm{n}} \pi^-$
or $\overline{\mrm{p}} \pi^0$;
\item $\overline{\Xi^-} \rightarrow \overline{\Lambda} \pi^+
\rightarrow \overline{\mrm{p}} \pi^+\pi^+$;
\item $\overline{\Xi^0} \rightarrow \overline{\Lambda} \pi^0
\rightarrow \overline{\mrm{p}} \pi^+ \pi^0$;
\item $\overline{\Lambda} \rightarrow \overline{\mrm{p}} \pi^+$,
including $\overline{\Lambda}$ antihyperons from 
$\overline{\Sigma^0}$ decays,
but without the contributions from $\overline{\Xi}$ decays;
\item charged particles from $\mrm{K}^+$ decays;
\item charged particles or V$^0$ configurations from $\mrm{K}^0$ decays;
\item non-$\mrm{K}^0$ background of the V$^0$ topology;
\item displaced tracks from secondary interactions, mainly 
scattering.
\end{enumerate}
All sources, except for the eighth one, contribute to the sample of
displaced tracks. For V$^0$-like events only the sources 
3,4,5,7 and 8 are relevant.
The sample of displaced tracks is enriched in pions but contains
also a small fraction of protons and leptons.
These contaminations are included in the Monte Carlo rates
and are classified according to the above scheme.
 
The combination of five kink sources with nine correlated particle 
sources leads to a total number of 45 classes of particle pairs.
In addition, one has to distinguish between like-sign and
unlike-sign pairs. 
Equation~(\ref{eq:kink-bin}) can be generalized to
the like-sign pair rate $D^{like}(j,d_0)$ and the
difference of the unlike-sign pair rate $D^{unlike}(j,d_0)$ 
and $D^{like}(j,d_0)$,
\begin{align}
\label{eq:corrd0}
D^{like}(j,d_0) &=  
\sum_{i,k}  \eta_{i,k}^{(l)} \cdot 
\frac{N_{data}}{N_{MC}}
\cdot D^{(MC,like)}_{i,k}(j,d_0) \ ;
\\
\label{eq:asyd0}
D^{unlike}(j,d_0)-D^{like}(j,d_0) &=  
\sum_{i,k}  \eta_{i,k}^{(a)} \cdot 
\frac{N_{data}}{N_{MC}}
\cdot (D^{(MC,unlike)}_{i,k}(j,d_0)-D^{(MC,like)}_{i,k}(j,d_0))   \ .
\end{align}
The indices at the Monte Carlo rates 
$D^{(MC,like)}_{i,k}$ and $D^{(MC,unlike)}_{i,k}$
specify the kink source $i$ and the correlated particle
source $k$, respectively. The parameters 
$\eta_{i,k}^{(l)}$ and $\eta_{i,k}^{(a)}$
are 90 scaling factors.
In principle, they have to be extracted with a combined fit of 
the $j$ and $d_0$ dependent distributions of kink-track
pairs and kink-V$^0$ pairs to equations~(\ref{eq:corrd0}) 
and~(\ref{eq:asyd0}).

\begin{table}
\begin{center}
\begin{tabular}{|c||c|c|c|c|c|c|}\hline
& \multicolumn{6}{c|}{correlated particle}\\ \cline{2-7}
\raisebox{1.5ex}[1.2\height]{kink source} &
\raisebox{0ex}[1.2\height]{$\overline{\Sigma^-}$} & 
\raisebox{0ex}[1.2\height]{$\overline{\Sigma^+}$} & 
\raisebox{0ex}[1.2\height]{$\overline{\Xi^-}$} &
\raisebox{0ex}[1.2\height]{$\overline{\Xi^0}$} & 
\raisebox{0ex}[1.2\height]{$\overline{\Lambda}$} & 
\raisebox{0ex}[1.2\height]{$\mrm{K}^+$}\\
\hline
\hline
$\Sigma^-$ & 0.33 &  0.014 & 0.15  & 0.01 & 0.28 & 0.04 \\ 
\hline 
$\Sigma^+$ & 0.01 & 0.32 & 0.01 &  0.14 & 0.28 & 0.024\\ 
\hline 
$\Xi^-$  & 0.34 &   0.026 & 0.15 & 0.029 & 0.36 & 0.45 \\ 
\hline 
$\mrm{K}^-$ & $<0.01$ & $<0.01$ & $<0.01$ & $<0.01$ & 0.03 & 0.50\\
\hline 
\end{tabular}
\end{center}
\caption{\sl \large Fractions $F^{(MC)}_{i,k}$ as predicted by the
PYTHIA 6.1 generator with the parameters from ref~\cite{opaltunejt74}.
}
\label{table:MCcorr}
\end{table}

The Monte Carlo generator predicts the generic 
correlations $F_{i,\overline{k}}^{(MC)}$, defined with
equation~(\ref{eq:pair-fraction}), generalized to arbitrary 
sources $i$ and $k$. 
Table \ref{table:MCcorr} gives the results for hyperon-antiparticle
correlations at generator level for the most recent parameter set
used by the OPAL experiment ~\cite{opaltunejt74}.

If the scaling factors $\eta_{i,k}^{(a)}$ are known, the
experimental results for the correlations can be computed 
\begin{eqnarray}
\label{eq:etaF}
F_{i,\overline{k}}= \frac{\eta_{i,k}^{(a)}}{\xi_i}
                    \cdot F_{i,\overline{k}}^{(MC)} \ .
\end{eqnarray}
Here, the scaling factors $\xi_i$ from the kink fit 
(section \ref{sect:analysis}.1) enter.

The factors $\eta_{i,k}^{(l)}$ are only needed to 
parameterize the statistical errors for the fit using 
equation~(\ref{eq:asyd0}).
With very few exceptions, they are close to one and no physical 
result is extracted from them.

\subsection{Evaluation of the correlation matrix}

In spite of the large number of scaling factors $\eta_{i,k}^{(a)}$,
a reliable fit of the data can be obtained.
This can be seen from the integrals of the differences 
$ (D^{(MC,unlike)}_{i,k}(j,d_0)-D^{(MC,like)}_{i,k}(j,d_0)) $
over the variables $m_{\Sigma^-},\cos \theta^*$ and $d_0$,
listed in tables ~\ref{table:MCcorrtrk}
and ~\ref{table:MCcorrv0}.
The entries in the tables are normalized to the 
total differences, using $\eta_{i,k}^{(a)}=1$, and
thus show the relative importance of the terms.
The nine most significant correlations are given in the tables,
the individual contributions of all other sources being less than 2\%.
In total, the entries in table~\ref{table:MCcorrtrk} for displaced tracks 
account for 96\% of the total rate difference. The contributions 
to the V$^0$ in table~\ref{table:MCcorrv0} add up to 104\%, 
the excess being compensated by a small amount of 
$\Sigma^+\overline{\Lambda}$ correlations with the opposite sign.

\begin{table}
\begin{center}
\begin{tabular}{|c||c|c|c|c|c|}\hline
& \multicolumn{5}{c|}{displaced track source}\\ \cline{2-6}
\raisebox{1.5ex}[1.2\height]{kink source} &
\raisebox{0ex}[1.2\height]{$\overline{\Sigma^-}$} & 
\raisebox{0ex}[1.2\height]{$\overline{\Xi^-}$} & 
\raisebox{0ex}[1.2\height]{$\overline{\Lambda}$} &
\raisebox{0ex}[1.2\height]{$\mrm{K}^+$} & 
\raisebox{0ex}{background} \\
\hline
\hline
$\Sigma^-$ & $(27 \pm 1)$\% & $(16 \pm 1)$\% & $(10 \pm 1)\%$ & & \\
\cline{1-5}
$\Xi^-$    & $(12 \pm 1)$\% & $(6 \pm 1)$\% & $(5 \pm 1)$\% & & \\
\cline{1-5}
$\mrm{K}^-$            &   &   &  & $(3 \pm 1)$\%  &
\raisebox{1.5ex}[0cm][0cm]{$(17 \pm 4)$\%} \\
\cline{1-5}
kink background & & & & & \\
\hline
\end{tabular}
\end{center}
\caption{\sl {\large Largest contributions to the correlations
between kinks and displaced tracks.
}}
\label{table:MCcorrtrk}
\end{table}

\begin{table}
\begin{center}
\begin{tabular}{|c||c|c|c|} \hline
& \multicolumn{3}{c|}{V$^0$ source}\\ \cline{2-4}
\raisebox{1.5ex}[0cm]{kink source} &
\raisebox{0ex}[1.2\height]{$\overline{\Xi^-}$} & 
\raisebox{0ex}[1.2\height]{$\overline{\Lambda}$} & 
\raisebox{0ex}[1.2\height]{V$^0$ background} \\
\hline
\hline
$\Sigma^-$ & $(20 \pm 2)$\% & $(43 \pm 3)$\% & 
\\ \cline{1-3}
$\Xi^-$    & $(6 \pm 1.5)$\% & $(21\pm 2)$\%  & 
\raisebox{1.5ex}[0cm][0cm]{$(14 \pm 3.5)$\%} \\ 
\hline
\end{tabular}
\end{center}
\caption{\sl {\large Largest contributions to the correlations
between kinks and V$^0$ candidates.
}}
\label{table:MCcorrv0}
\end{table}

The scaling factors for the four largest correlations in 
tables~\ref{table:MCcorrtrk} and \ref{table:MCcorrv0},
$\eta_{\Sigma^-,\overline{\Sigma^-}}^{(a)}$,
$\eta_{\Sigma^-,\overline{\Lambda}}^{(a)}$,
$\eta_{\Sigma^-,\overline{\Xi^-}}^{(a)}$ and 
$\eta_{\Xi^-,\overline{\Lambda}}^{(a)}$,
are determined with the fit. Three of them are needed to 
compute $F_{\overline{H}}$.

The data statistics do not allow to fit more than four parameters.
The remaining $\eta$'s were thus fixed by
symmetry considerations, isospin invariance or
Monte Carlo studies. Systematic errors were assigned to them, if 
necessary. In the following, a few examples are described, 
preferentially the correlations in 
tables~\ref{table:MCcorrtrk},\ref{table:MCcorrv0}.

For symmetry reasons, one has 
\begin{eqnarray}
\eta_{\Xi^-,\overline{\Sigma^-}}^{(a)}=
\eta_{\Sigma^-,\overline{\Xi^-}}^{(a)} \ .
\end{eqnarray}
There is no model independent prediction for $F_{\Xi^-,\overline{\Xi^-}}$.
This parameter was set to the original Monte Carlo result. 
It follows then from equation~(\ref{eq:etaF}) that
\begin{eqnarray}
\label{eq:modelconstr}
\eta^{(a)}_{\Xi^-,\overline{\Xi^-}} = \xi_{\Xi^-}  \ .
\end{eqnarray}
It has been checked for various acceptable Monte Carlo generator 
tunings, described in the next section, that this procedure
is valid within 20\%.

Non-negligible parts of the correlation are introduced by the
correlated particle backgrounds 8 and 9; they are given as 
sums over the kink sources in tables~\ref{table:MCcorrtrk} 
and~\ref{table:MCcorrv0}.
The origin of this effect is charge conservation in the events.
Since the displaced-track-background consists mainly of scattered 
particles, it reflects the original particle charges. The kink 
selection introduces a charge bias for the rest of the event, which is 
visible in the ensemble of remaining charged particles on a statistical 
basis. This is true not only for scattered particles, but also for
asymmetric fake V$^0$ candidates. Kinks and correlated particles 
are assigned to each other by chance. The condition
$\eta^{(a)}_{i,track}=\eta^{(a)}_{i,V^0}=\xi_i$
was introduced, assuming a correct modeling of the background 
sources 8 and 9. This is justified because
the equivalent relation for the like-sign background rates was 
confirmed with a fit to the data with the parameterizations
$\eta^{(l)}_{i,track}=\xi_i \zeta_{track}$ and
$\eta^{(l)}_{i,V^0}=\xi_i \zeta_{V^0}$ , the results being
$\zeta_{track}= 1.01 \pm 0.04$ and
$\zeta_{V^0}=1.03 \pm 0.07$, respectively.

The remaining entry in table \ref{table:MCcorrtrk} is the charged 
kaon-kaon correlation.
Its contribution is small because kaon decay inside the jet chamber is
unlikely due to the long kaon life time. Since the charged and
neutral kaon production rates are almost equal and the hyperon
rates are much smaller than the kaon rates,
the fraction $F_{K^-,K^+}$ is close to 0.5 so that 
$\eta^{(a)}_{K^-,K^+}=\xi_{K^-}$.

In total, the contribution of all other sources to the observed 
correlation is smaller than the statistical error of the final
result. Nevertheless, all sources were investigated in detail to
minimize the systematic error. 

A correlation potentially dangerous for the fit 
is the $\Sigma^-\mrm{K}^+$ correlation. The corresponding
asymmetry in table~\ref{table:MCcorr} is small, but actually it is the
difference of much larger components. Strangeness conservation 
requires
\begin{eqnarray}
\label{eq:Kconstraint}
  F_{\Sigma^-,\overline{\Sigma^-}}
+ F_{\Sigma^-,\overline{\Sigma^+}}
+ 2 \cdot F_{\Sigma^-,\overline{\Xi^-}}
+ 2 \cdot F_{\Sigma^-,\overline{\Xi^0}}
+ F_{\Sigma^-,\overline{\Lambda}}
+ 2 \cdot F_{\Sigma^-,K^+} \approx 1 \ .
\end{eqnarray}
The parameters $F_{\Sigma^-,\overline{\Xi^-}}$ and
$F_{\Sigma^-,\overline{\Xi^0}}$ appear with the weight 2 because
the $\Xi$-particles carry two units of strangeness.
The contribution from $\overline{\Omega^-}$ hyperons is negligible. 
The factor 2 in front of $F_{\Sigma^-,K^+}$ accounts for 
the $\overline{\mrm{K}^0}$ contribution, which cannot be measured.
Monte Carlo simulations with different model parameters showed
that relation (\ref{eq:Kconstraint}) is fulfilled within 3\%.
Together with the condition for baryon number conservation,
\begin{eqnarray}
 F_{\Sigma^-,\overline{\Sigma^-}}+F_{\Sigma^-,\overline{\Sigma^+}}
+F_{\Sigma^-,\overline{\Xi^-}}   +F_{\Sigma^-,\overline{\Xi^0}}
+F_{\Sigma^-,\overline{\Lambda}}
+F_{\Sigma^-,\overline{p}}+F_{\Sigma^-,\overline{n}} \approx 1,
\end{eqnarray}
equation (\ref{eq:Kconstraint}) gives the relation
\begin{eqnarray}
 F_{\Sigma^-,K^+} \approx \frac{1}{2}\cdot (
 F_{\Sigma^-,\overline{p}}     + F_{\Sigma^-,\overline{n}}
-F_{\Sigma^-,\overline{\Xi^-}} - F_{\Sigma^-,\overline{\Xi^0}} )\ ,
\end{eqnarray}
which shows the presence of large compensating terms.
In the analysis, $F_{\Sigma^-,K^+}$ was therefore computed with the
sum rule~(\ref{eq:Kconstraint}), taking the small correlations 
$F_{\Sigma^-,\overline{\Sigma^+}}$ and $F_{\Sigma^-,\overline{\Xi^0}}$
from the Monte Carlo generator.
Similarly, $F_{\Xi^-,K^+}$ can be 
constrained by the equivalent equation for the $\Xi^-$,
\begin{eqnarray}
\label{eq:KXconstraint}
  F_{\Xi^-,\overline{\Sigma^-}}
+ F_{\Xi^-,\overline{\Sigma^+}}
+ 2 \cdot F_{\Xi^-,\overline{\Xi^-}}
+ 2 \cdot F_{\Xi^-,\overline{\Xi^0}}
+ F_{\Xi^-,\overline{\Lambda}}
+ 2 \cdot F_{\Xi^-,K^+} \approx 2 \ .
\end{eqnarray}
The correlations of the $\Sigma^+$ hyperons 
are related to those of the $\Sigma^-$ hyperon by the isospin
symmetry, for instance 
$F_{\Sigma^+,\overline{\Lambda}}=F_{\Sigma^-,\overline{\Lambda}}$.
No model independent predictions exist for $F_{K^-,\overline{\Lambda}}$
and the popcorn specific correlations 
$F_{\Sigma^-,\overline{\Xi^0}}$ and $F_{\Sigma^-,\overline{\Sigma^+}}$,
which were determined with the Monte Carlo event sample.

The above relations allow either to replace $\eta_{i,k}^{(a)}$
by the four factors to be fitted, or to
fix it and its contribution to the right hand side 
of equation~(\ref{eq:asyd0}).
A simultaneous binned $\chi^2$ fit, using 
equation~(\ref{eq:asyd0}), was performed 
for the $d_0$ distributions of the correlated-track and $\Lambda$ 
candidates in the five $(m_{\Sigma^-}, \cos \theta^*)$-regions. 
In the $\Lambda$ case, the reconstructed mass was restricted to the 
narrow interval given in section~\ref{sect:lambdaselection}.
In parallel, a fit of the like sign pairs
to equation~(\ref{eq:corrd0}) was performed to determine
$\zeta_{track} = \eta^{(l)}_{i,track} / \xi_i$ and three 
normalization factors for the hyperon-$\overline{\Lambda}$, 
K$^-$-hyperon and ($\Sigma^-$ or $\Xi^-)-(\overline{\Sigma^-}$ 
or $\overline{\Xi^-})$ correlations.
The statistical errors of the pair rates, needed to compute $\chi^2$,
were computed with the equations~(\ref{eq:corrd0}) and~(\ref{eq:asyd0}); 
they depend on the result of the fit.
The fit was therefore done iteratively, setting the unknown 
$\eta$-factors to unity for the first iteration.
The statistical errors had $d_0$ dependent fluctuations due
to the limited Monte Carlo statistics. These were reduced
with a one-dimensional smoothing algorithm described in~\cite{smoothing}.
The asymmetry fit resulted in a $\chi^2$ value of 144 for 146 
degrees of freedom. 
The coefficients obtained are correlated, the largest correlation 
coefficient being $\approx$-0.7 between
$\eta^{(a)}_{\Sigma^-,\overline{\Sigma^-}}$ and
$\eta^{(a)}_{\Sigma^-,\overline{\Xi^-}}$.

\subsection{Experimental results and systematic errors}

\begin{figure}[htb] 
\centering
\includegraphics[width=0.9\textwidth]{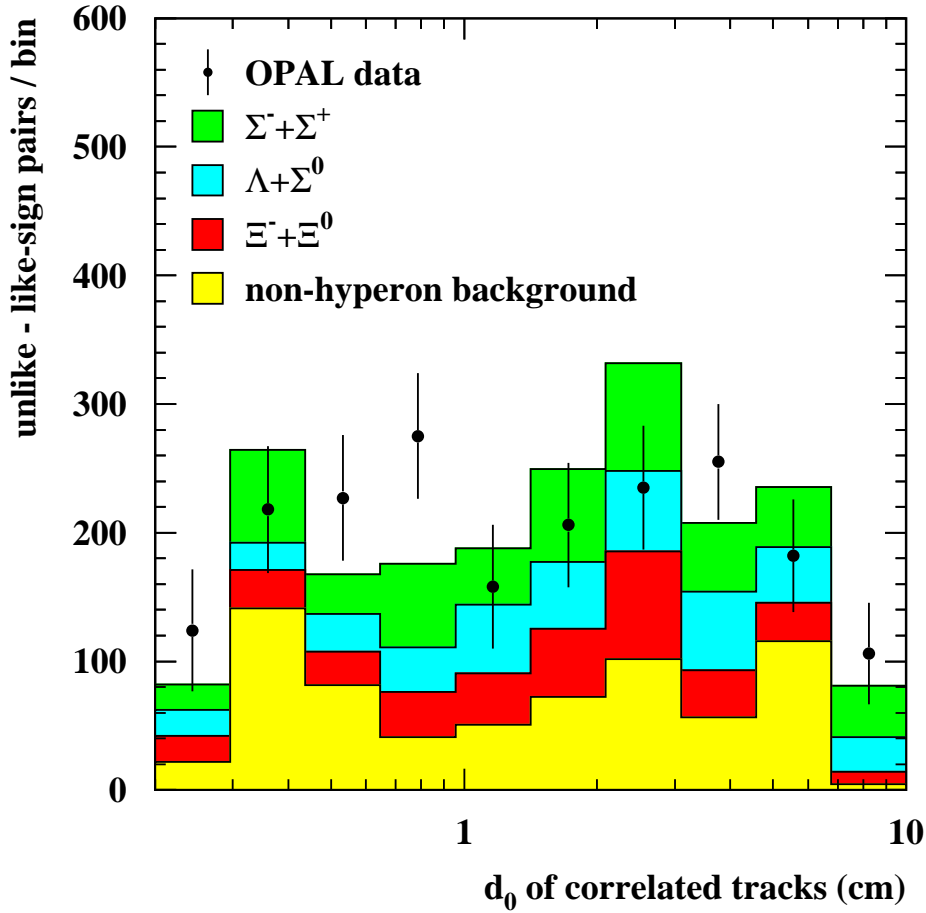}
\caption{ \sl Impact parameters of the displaced tracks,
correlated to $\Sigma^-$ candidates.
The difference between unlike-sign and like-sign 
combinations is shown. Points: data. Histograms: results of the fit, 
ordered according to the sources for displaced tracks.
The plotted errors are the statistical errors of the data.}
\label{fig:d0-asy-track}
\end{figure}

\begin{figure}[htb] 
\centering
\includegraphics[width=0.9\textwidth]{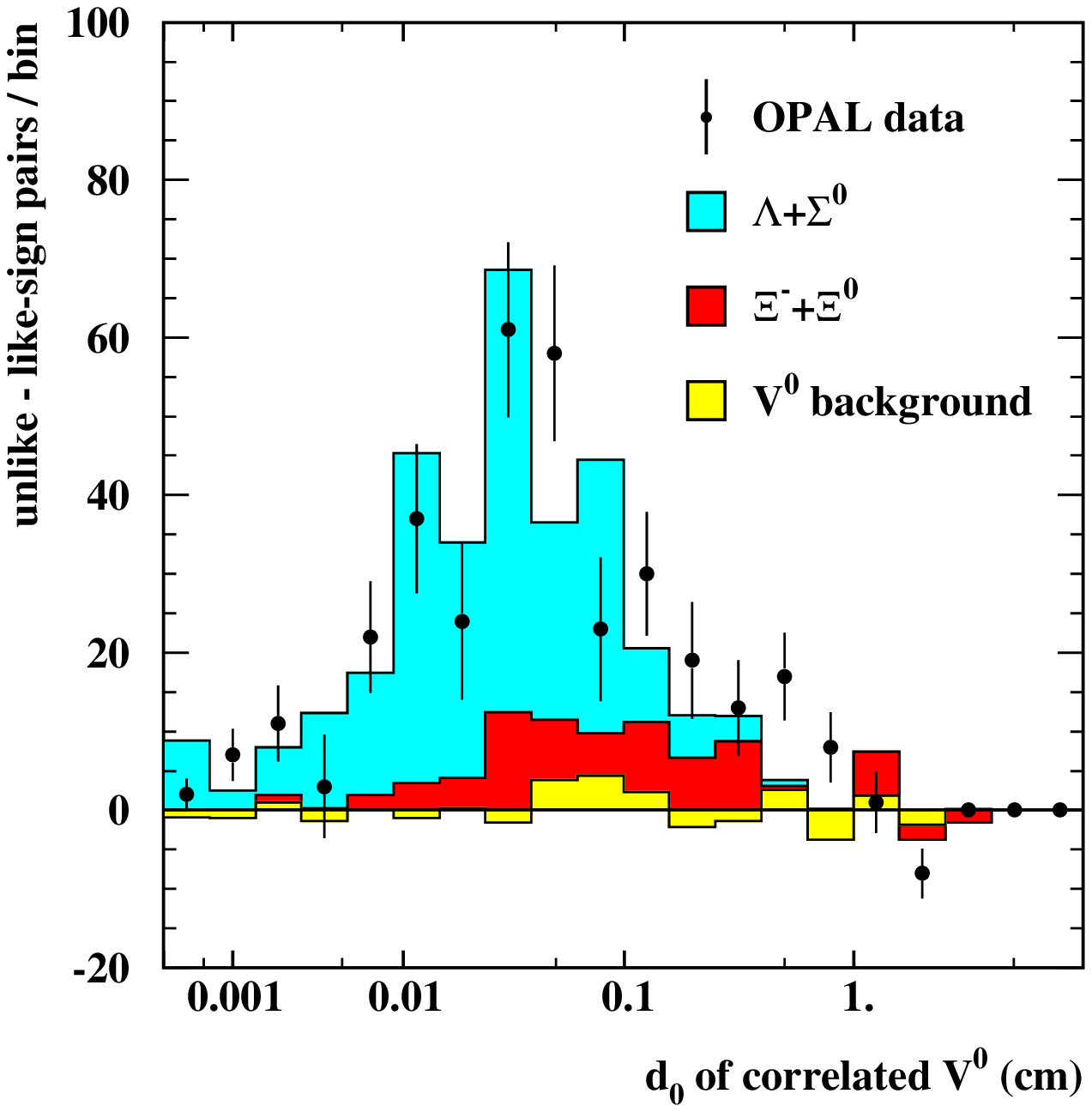}
\caption{ \sl Impact parameters of the $\overline{\Lambda}$ candidates, 
correlated to $\Sigma^-$ candidates.
The difference between unlike-sign and like-sign 
combinations is shown. Points: data. Histograms: results of the fit,
ordered according to the V$^0$ sources. The plotted errors are the 
statistical errors of the data.}
\label{fig:d0-asy-v0}
\end{figure}

\begin{figure}[htb] 
\centering
\includegraphics[width=0.9\textwidth]{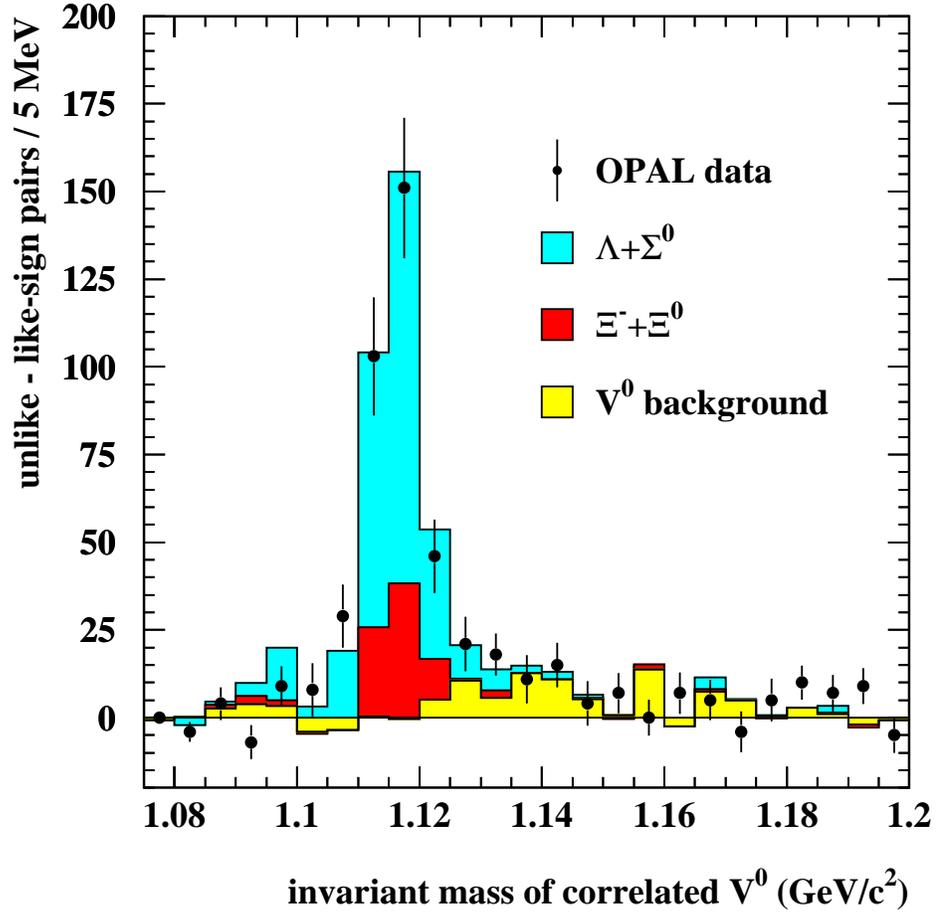}
\caption{ \sl Reconstructed masses of the $\overline{\Lambda}$ candidates, 
correlated to $\Sigma^-$ candidates. 
The difference between unlike-sign and like-sign combinations is shown. 
Points: data. Histograms: results of the fit,
ordered according to the V$^0$ sources. The plotted errors are the 
statistical errors of the data.}
\label{fig:mass-asy-v0}
\end{figure}

The adjusted $d_0$ distributions are shown in 
figs.~\ref{fig:d0-asy-track} and \ref{fig:d0-asy-v0}.
The histograms give the contributions of the correlated particle 
sources; all kink sources and two-dimensional bins $j=${\it a,...e} 
are combined. 
The errors of the data points are statistical and the corresponding errors 
of the Monte Carlo histograms are not shown.
Fig. \ref{fig:mass-asy-v0} gives the rate differences between the
unlike-sign and like-sign kink-V$^0$ pairs
as a function of the reconstructed $\Lambda$ mass.
Both distributions are very well described.
\clearpage
\clearpage

The fit results are listed in the first row of table \ref{table:fitresult}.
The errors are statistical and include the data and Monte Carlo 
contributions. 
Internally, the fit gets the $\Lambda$ part of the correlation 
essentially from the correlated $\Lambda$ sample, which sets also 
bounds on $F_{\Sigma^-,\overline{\Xi^-}}$.
The $\Sigma^- \overline{\Sigma^-}$ correlation is computed
from the displaced-track sample as a difference. Neither
$F_{\Sigma^-,\overline{\Sigma^-}}$ nor $F_{\Sigma^-,\overline{\Xi^-}}$ 
differ from zero in a statistical way, in contrast to the overall 
sum $F_{\overline{H}}$. The smaller  error of the sum is due to the 
strong anti-correlation between $F_{\Sigma^- \overline{\Sigma^-}}$ 
and $F_{\Sigma^-,\overline{\Xi^-}}$.
The final results, including the systematic errors, are given in the last
row of table \ref{table:fitresult}. \\

\begin{table} [t]
\begin{center}
\begin{tabular}{|c||c|c|c|c|}\hline
 & \multicolumn{4}{c|}{correlation}\\ \cline{2-5}
 & $F_{\Sigma^-,\overline{\Sigma^-}}$ 
           & $F_{\Sigma^-,\overline{\Xi^-}}  $ 
           & $F_{\Sigma^-,\overline{\Lambda}} $
           & $F_{\overline{H}}                $ \\
\hline
fit result & $0.162 \pm 0.100$ & $0.054 \pm 0.047$ & $0.234 \pm 0.057$  
           & $0.449 \pm 0.091$ \\
\hline
\hline
systematic error & \multicolumn{4}{c|}{}\\ \cline{2-5}
\hline 
mismodeling of correlated & +0.009 & +0.003  & +0.014 & +0.026 \\ 
particle momenta & $\pm 0.005$ & $\pm 0.002$ & $\pm 0.007$ &
$\pm 0.013$ \\
\hline
$z$-resolution at  & & & & \\
kink vertex      &
\raisebox{1.5ex}[0cm][0cm]{$\pm 0.037$} &
\raisebox{1.5ex}[0cm][0cm]{$\pm 0.023$} &
\raisebox{1.5ex}[0cm][0cm]{$\pm 0.045$} &
\raisebox{1.5ex}[0cm][0cm]{$\pm 0.024$} \\
\hline
$dE/dx$ calibration & $\pm 0.010$ & $\pm 0.005$ & $\pm 0.004$ & $\pm 0.008$ \\
\hline
hemisphere cut & $\pm 0.005$ & $\pm 0.002$ & 
                 $\pm 0.007$ & $\pm 0.013$ \\
\hline
number of hits per track & $\pm 0.007$ & $\pm 0.004 $ & $\pm 0.002$ 
                         & $\pm 0.006$ \\
\hline 
$\cos \theta$ distribution of tracks & $\pm 0.012$ & $\pm 0.001$ & 
                                       $\pm 0.001$ & $\pm 0.012$ \\
\hline
$\Lambda$ detection efficiency & $\pm 0.006$ & $\pm 0.004$ & 
                                $\pm 0.004$ & $\pm 0.002$ \\
\hline
charge asymmetry          &  &  &  &  \\         
of detection efficiencies & 
\raisebox{1.5ex}[0cm][0cm]{$\pm 0.002$} &
\raisebox{1.5ex}[0cm][0cm]{$\pm 0.001$} &
\raisebox{1.5ex}[0cm][0cm]{$\pm 0.001$} &
\raisebox{1.5ex}[0cm][0cm]{$\pm 0.002$} \\
\hline
uncertainty of                    & & & & \\
popcorn channels                  & 
\raisebox{1.5ex}[0cm][0cm]{$\pm 0.008$} & 
\raisebox{1.5ex}[0cm][0cm]{$\pm 0.005$} & 
\raisebox{1.5ex}[0cm][0cm]{$\pm 0.004$} & 
\raisebox{1.5ex}[0cm][0cm]{$\pm 0.007$} \\
\hline
uncertainty of                    & & & & \\
$\Xi^- \overline{\Xi^-}$ correlation & 
\raisebox{1.5ex}[0cm][0cm]{$\pm 0.003$} & 
\raisebox{1.5ex}[0cm][0cm]{$\pm 0.001$} & 
\raisebox{1.5ex}[0cm][0cm]{$\pm 0.001$} & 
\raisebox{1.5ex}[0cm][0cm]{$\pm 0.003$} \\
\hline
uncertainty of                    & & & & \\
$\mrm{K}^-\overline{\Lambda}$ correlation & 
\raisebox{1.5ex}[0cm][0cm]{$\pm 0.001$} & 
\raisebox{1.5ex}[0cm][0cm]{$\pm 0.001$} &
\raisebox{1.5ex}[0cm][0cm]{$\pm 0.003$} & 
\raisebox{1.5ex}[0cm][0cm]{$\pm 0.002$}\\
\hline
\hline
final result & $ 0.17 \pm 0.11 $ & $ 0.057 \pm 0.056 $
             & $ 0.25 \pm 0.08 $ & $ 0.48  \pm 0.10 $ \\
\hline
\end{tabular}
\end{center}
\caption{\sl \large Experimental results and systematic errors.
}
\label{table:fitresult}
\end{table}

The systematic errors, listed in
table \ref{table:fitresult}, will be discussed in the following.
\begin{description}
\item[{Mismodeling of correlated particle momenta.}]
{The momentum spectra of the correlated pion and 
$\Lambda$ candidates are not well reproduced by the Monte Carlo 
program. The simulated momenta have to be scaled downwards by
as much as 20\%. The effect exists both for the like sign and the 
unlike-sign track and V$^0$ candidates. As a consequence, the measured 
fractions $F_{i,\overline{k}}$ are systematically too small. To find a 
correction, the detection efficiencies for the correlated particles 
were extracted from the 
Monte Carlo sample and the shifted momentum spectrum was folded 
with the efficiency function. This leads to an upwards correction 
of the measured correlations. A global correction of $(6 \pm 3)$\% 
was applied.
This correction, together with its error, is much larger than the 
uncertainties due to miscalibrations of the particle momenta or 
the impact parameters. Therefore, no additional
errors were assigned to the momentum and $d_0$ selection cuts.}
\item[{$z$-resolution at kink vertex.}]
{In the ratio of a correlated rate to the single $\Sigma^-$ rate the 
overall detection efficiency for $\Sigma^-$ hyperons cancels.
However, an uncertainty of the number of observed
$\Sigma^-$ hyperons arises from the unfolding procedure.
The dominating error source is the mismodeling of the $z$-resolution.
The correction factor $c$, introduced above, was conservatively varied 
between 1.25 and 1.5. For the individual correlations 
$F_{\Sigma^-,\overline{k}}$ 
the obtained shifts were found to be non-parabolic functions of $c$.
The maximal shifts of the results $F_{\Sigma^-,\overline{k}}$ and 
$F_{\overline{H}}$ were taken as uncertainties.}
\item[{$dE/dx$ calibration.}]
{Any antiproton impurity in the displaced track sample reduces the 
asymmetry from direct or indirect $\overline{\Lambda}$ decays.
For a pure pion sample and a perfect calibration, the frequency 
distribution of the weight $w_{dE/dx}(\pi)$, defined in 
section~\ref{sect:pionselection}, should not depend on $w_{dE/dx}(\pi)$.
A superimposed peak at $w_{dE/dx}(\pi)=0$ is expected due 
to non-pions. The shape of this peak does not perfectly agree with the 
Monte Carlo prediction. A systematic error was assigned to the
corresponding mismodeling of the antiproton rejection efficiency. 
}
\item[{Hemisphere cut.}]
{In principle, the results are corrected automatically for the 
hemisphere cut. No significant correlations were observed in the 
dropped hemisphere. A systematic error would appear, if the fragmentation
model were incorrect. A 3\% error was assigned to all correlations,
based on the error of the number of tracks in the omitted hemisphere.}  
\item[{Other systematic errors}]
{The number of hits per track is well modeled and introduces an 
error of 1\% for the number of displaced tracks.} 
{Also the acceptance
cut for the $\theta$ angle plays a minor role only. From the difference
between the angular distributions of data and Monte Carlo events an
efficiency error of 3\% was estimated.}
{Systematic errors due to the cuts for $\Lambda$ selection were
already discussed in~\cite{strange-baryons-opal}.
An uncertainty of $\pm 3.3\%$ was taken from that paper
as fully correlated error for the $\Lambda$ event sample.}
{Differences $\Delta \epsilon_i$ between the detection efficiencies 
for particles and antiparticles would lead to a spurious asymmetry,
if both the kinks $i$ and the correlated particles $k$ are affected.
The effect is proportional to 
$\Delta \epsilon_{\Sigma^-} \cdot \Delta \epsilon_k$.
The efficiency differences were extracted from the observed 
particle and antiparticle rates and the upper limits for the spurious 
asymmetries in table~\ref{table:fitresult} were obtained.
} 
{Furthermore, there are uncertainties due to the model dependence of
the correlations which had to be subtracted.
One half of the Monte Carlo baryon antibaryon pairs are accompanied 
by a popcorn meson. 
The $\eta^{(a)}$-factors for the popcorn specific
correlations $\Sigma^- \overline{\Sigma^+}$, 
$\Sigma^- \overline{\Xi^0}$ and $\Xi^- \overline{\Sigma^+}$ 
were varied between 0 and 2 and the fit repeated.}
{The $\mrm{K}^- \overline{\Lambda}$ correlation
was varied by 50\% and an uncertainty of 20\% was assigned to 
the $\Xi^- \overline{\Xi^-}$ correlation, fixing the
small residual contribution from $\Xi^-$ self correlations.} 
{As already mentioned, the definition of $F_{\overline{H}}$ introduces
a small amount of double counting, both in the data and in the 
Monte Carlo simulation. 
This problem was studied with a toy Monte Carlo program and the 
only effects found were
negligible corrections to the statistical errors of the fit.}
\end{description}

The correlated kink-track and kink-V$^0$ pairs
should be concentrated in the $\Sigma^-$ enriched bins $a$ and $d$
in the $(m_{\Sigma^-}, \cos \theta^*)$ plane shown 
in figure~\ref{fig:mass-vs-cost}.
As a cross check, a simpler analysis was performed, where
the analysis was restricted to region $a$.
All background asymmetries and also the $\Xi^-\overline{\Lambda}$ 
correlation were subtracted as predicted by the 
Monte Carlo program. The final result including reevaluated systematic 
errors, $F_{\overline{H}}= 0.472 \pm 0.155$, is fully consistent with 
the main result and demonstrates the absence of anomalies in the  
$(m_{\Sigma^-},\cos \theta^*)$ plane.

\section{Comparison with fragmentation models}
\label{sect:interpretation}

Before comparing the result for $F_{\overline{H}}$ to the
predictions of models, we adjusted the models, incorporated into 
the PYTHIA Monte Carlo event generator,
to describe a set of observables in \Zo~decays. 
The optimization began with the
PYTHIA steering parameters given in~\cite{opaltunejt74}.
These were slightly modified to reproduce the newest experimental 
information on the baryonic sector, including data on
$\Lambda \overline{\Lambda}$ correlations.
The tuned Monte Carlo was then used to predict $F_{\overline{H}}$. 
We also used the models to study 
$\mrm{p} \pi \overline{\mrm{p}}$ correlations more extensively than 
previously, as described below.

\subsection{Observables}
\label{sect:observables}
 The input observables for the tuning fall into four categories:
\begin{enumerate}
\item {Eight baryon multiplicities in multihadronic \Zo~decay:}  
$\mrm{p}$, $\Delta^{++}$, $\Lambda$, $\Sigma^+ + \Sigma^-$, $\Sigma^0$, 
$\Xi^-$, $\Sigma^{*+}+\Sigma^{*-}$, and $\Omega^-$.
The production rates were taken from the compilation of 
the particle data group~\cite{PDGmanual}.

\item {Proton and $\Lambda$ momentum spectra.} 
Baryon spectra have approximately Gaussian shapes if 
parameterized in terms of the variable $\xi=\ln(1/x_p)$ ~\cite{lnxp}. 
We use the mean values and variances of the proton and $\Lambda$
spectra as the observables.
Deviations from the normal distribution 
are known, the true \mbox{maxima} being somewhat higher than the result of 
the fit~\cite{proton-delphi,rates-SLD}. However, the Gaussian fit is 
an easy way to compare different experiments. 

The tabulated data ~\cite{proton-delphi}--\cite{rates-SLD} 
were fitted to Gaussian functions in the interval $1.2 < \xi < 4.2$. 
The situation for the protons is not satisfactory,
the fitted maxima of the $\xi$ distributions varying from 
2.79~\cite{proton-aleph} to 3.08~\cite{rates-SLD}.
  The values $\xi_{peak}=2.80 \pm 0.07$ and $\sigma=1.11 \pm 0.06$
were obtained with a combined fit using all LEP and SLD data.
The errors were not taken from the fit but conservatively estimated 
from the systematic differences between the $\xi$ spectra
of the experiments. It should be noted 
that both $\xi_{peak}$ and $\sigma$ depend on the fit range 
due to deviations of the $\xi$-distribution from the Gaussian shape.
The agreement between experiments is better for $\Lambda$
 production, leading to values $\xi_{peak}= 2.62 \pm 0.04$
 and $\sigma=1.21 \pm 0.04 $.

\item {$\Lambda \overline{\Lambda}$ correlations.}
The correlation was parameterized by two observables,
the first one being the rate excess of $\Lambda \; \overline{\Lambda}$ pairs 
over $\Lambda \; \Lambda\; +\; \overline{\Lambda} \;\overline{\Lambda}$ pairs
per event, $N^{corr}_{\Lambda \overline{\Lambda}}$. The other observable
is the mean rapidity difference 
$\overline{\Delta y}_{\Lambda \overline{\Lambda}}$
between the $\Lambda$ and $\overline{\Lambda}$ after subtraction 
of like-sign pairs.
The number of correlated $\Lambda \; \overline{\Lambda}$ pairs per event 
was taken from~\cite{lambda-pair-opal}.
The mean $\Lambda\overline{\Lambda}$ rapidity difference 
$\overline{\Delta y}$ was computed from the data 
of~\cite{lambda-pair-opal}, as the
truncated mean for $\Delta y < 3.0$. This cut was introduced to suppress
the contribution of $\Lambda$'s from the opposite event hemisphere. 

\item {Nine meson production rates per multihadronic \Zo~decay:}
$ \pi^+ + \pi^- $, $ \pi^0 $, $ \mrm{K}^+ + \mrm{K}^- $, $\mrm{K}^0_S$, 
$\rho^+ + \rho^-$, $\rho^0$, $\varphi$, 
$\mrm{K}^{*+}(892) + \mrm{K}^{*-}(892) $ and $\mrm{K}^{*0}(892)$.
These rates, taken from~\cite{PDGmanual}, 
were included in the tuning to protect the meson generation against
parameter modifications steering the baryonic sector.
\end{enumerate}

\subsection{Model Parameters}
\label{sect:modelpars}
The popcorn mechanism is incorporated in PYTHIA 
in two ways ~\cite{pythia61}. The first {\bf simple version} was 
originally introduced in JETSET. Baryon production is controlled
by the following parameters:
\begin{enumerate}
\item the suppression of diquark-antidiquark
production relative to quark-antiquark production, 
${\mathbf{P}}(qq)/{\mathbf{P}}(q)=$ PARJ(1); 
\item the suppression of $s\overline{s}$ 
production relative to $u\overline{u}$ production,
${\mathbf{P}}(s)/{\mathbf{P}}(u)=$ PARJ(2); a tuning of this
parameter was necessary, because the strange meson rates had to be
readjusted;
\item a double ratio involving diquarks containing $s$ quarks,
$({\mathbf{P}}(us)/{\mathbf{P}}(ud))/
 ({\mathbf{P}}(s)/{\mathbf{P}}(u))=$ PARJ(3);
\item the suppression factor for spin 1 diquarks,
$(1/3){\mathbf{P}}(ud_1)/{\mathbf{P}}(ud_0)=$ PARJ(4);
\item the popcorn parameter, which determines the relative occurrences 
of the baryon-meson-antibaryon and baryon-antibaryon configurations, PARJ(5);
\item an extra suppression for having an $s\;\overline{s}$ pair in
a baryon-meson-antibaryon configuration, PARJ(6);
\item an extra suppression for having a strange meson in a
baryon-meson-antibaryon configuration, PARJ(7);
\item a parameter which enters the exponent of the Lund symmetric 
fragmentation function for diquarks, PARJ(45);
this parameter has an impact on the rapidity difference in 
baryon-antibaryon correlations.
\end{enumerate}
 
In the {\bf advanced popcorn} scheme, a universal equation for 
tunneling from the vacuum is
applied to the generation of new partons
and an arbitrary number of mesons can be created between a baryon 
and an antibaryon.
The tunneling probability is proportional to
\mbox{$\exp(-\beta_q \cdot M_{\perp})$},
where $\beta_q$ is a flavor dependent model parameter and $M_{\perp}$ the
transverse mass of the created object.
Only the first two and the last parameter of the above list  
are used in this scheme. There are three new parameters, 
two of them related to the tunneling formula:
\begin{enumerate}
\setcounter{enumi}{8}
\item the tunneling coefficient for u-quarks, $\beta_u=$ PARJ(8);
\item $\delta \beta=\beta_s-\beta_u$ = PARJ(9)
\item an extra suppression factor for spin 3/2 baryons = PARJ(18).
\end{enumerate}

\begin{table}
\begin{center}
\begin{tabular}{|c|c|r|r|r|r|c|}
\hline
 popcorn parameter   &    &  &  &  &  &  \\
 PARJ(5) &    & \raisebox{1.5ex}[0cm][0cm]{0.} &
    \raisebox{1.5ex}[0cm][0cm]{0.5}  &
    \raisebox{1.5ex}[0cm][0cm]{1.}   &
    \raisebox{1.5ex}[0cm][0cm]{5.}   & \\
\hline
\hline
            &       &      &      &      &     & MC5 \\
 observable & data   & MC1  & MC2  & MC3 & MC4 & advanced \\
            &       &      &      &      &     & popcorn  \\
\hline
$\chi^2$    &   -   &  56  &  59  & 59   &  63  & 99 \\
\hline
\raisebox{0ex}[1.4\height]{$N^{corr}_{\Lambda \overline{\Lambda}}$}
                   & $0.0612 \pm 0.0034$ 
                   & 0.066 & 0.060 & 0.066 & 0.058 & 0.081 \\
\raisebox{0.3ex}[1.5\height]
                   {$\overline{\Delta y}_{\Lambda \overline{\Lambda}}$}
                   & \raisebox{0.3ex}[1.5\height]{$0.71 \pm 0.04$}   
                   & \raisebox{0.3ex}[1.5\height]{0.66}  
                   & \raisebox{0.3ex}[1.5\height]{0.69}  
                   & \raisebox{0.3ex}[1.5\height]{0.67} 
                   & \raisebox{0.3ex}[1.5\height]{0.75} 
                   & \raisebox{0.3ex}[1.5\height]{0.57}\\
\hline
\hline
 $F_{\Sigma^-,\overline{\Sigma^-}}$ (this work) & $0.17 \pm 0.11$
                & 0.39 & 0.34 & 0.30 & 0.19 & 0.20 \\       
 $F_{\Sigma^-,\overline{\Xi^-}}$ (this work) & $0.057 \pm 0.056$ 
                & 0.18 & 0.16 & 0.15 & 0.09 & 0.08 \\ 
 $F_{\Sigma^-,\overline{\Lambda}}$ (this work)  & $0.25 \pm 0.08$ 
                & 0.30 & 0.28 & 0.28 & 0.27 & 0.28 \\ 
\hline
 \raisebox{0.3ex}[1.4\height]{$F_{\overline{H}}$ (this work)} 
                & \raisebox{0.2ex}[1.4\height]{$0.48 \pm 0.10$} 
                & \raisebox{0.2ex}[1.4\height]{0.87} 
                & \raisebox{0.2ex}[1.4\height]{0.79} 
                & \raisebox{0.2ex}[1.4\height]{0.73} 
                & \raisebox{0.2ex}[1.4\height]{0.55} 
                & \raisebox{0.2ex}[1.4\height]{0.56} \\
\hline
\end{tabular}
\end{center}
\caption{\it Comparison with fragmentation models.
Fit quality $\chi^2$, 
$\Lambda \overline{\Lambda}$ correlations and the
$\Sigma^-$ antihyperon correlations as defined in the text. The
errors of the simulation are smaller than the last digit shown.}
\label{table:MCcorrhyp}
\end{table}

\subsection {Simulation Results}

For a given PYTHIA parameter set, the measured observables
from section~\ref{sect:interpretation}.1 were compared with the
simulation results and a $\chi^2$ was computed.
It cannot be expected from a fragmentation model that all its
predictions are correct to better than a few percent.
To reduce the contributions of very accurately 
measured observables, the errors to compute $\chi^2$ were thus
taken to be at least 2.5\%, which represents the characteristic
level of agreement between the data and MC.
Low $\chi^2$ values were searched for with the method described 
in the appendix, for fixed values of the popcorn parameter. 
Many tunes of the generator have almost the same 
quality. Some PYTHIA parameter sets and the predicted
baryon production rates are given in the appendix. 
Table \ref{table:MCcorrhyp} shows the $\chi^2$'s, the 
$\Lambda \overline{\Lambda}$ correlations and the
$\Sigma^-$-antihyperon correlations, as a function of the
popcorn parameter, where the simple popcorn model is
denoted MC1 to MC4.

The variable $\chi^2$ is an indicator for the quality
of the baryon modeling. For the simple 
popcorn model, the lowest $\chi^2$ value found was
56 for the 23 observables. 
This means that the Monte Carlo generator 
describes, on average, the observables roughly within 2 times the 
experimental errors or 5\%, whichever is larger.

The most important result of the simulation is that the overall 
quality of the description of the observables does not depend 
strongly on the popcorn parameter in the simple popcorn model.
The $\Lambda \overline{\Lambda}$ correlation 
parameters are always reproduced within two standard deviations, 
whether the popcorn effect is switched on or off. 
The Monte Carlo parameter space examined here is larger than that
in earlier studies. In view of the overall uncertainty
it is not possible to reach a definite conclusion about the 
the popcorn effect by using $\Lambda \overline{\Lambda}$~correlations.

As shown in the bottom line of table \ref{table:MCcorrhyp},
the measured fraction $F_{\overline{H}}$ is consistent both with
the predictions of the original popcorn model with a large 
popcorn parameter (MC4) and the optimized advanced popcorn 
model~(MC5), but is smaller than the Monte Carlo prediction 
for zero popcorn effect. It was investigated whether models
without the popcorn mechanism could be found that reproduce
the observables of section~\ref{sect:interpretation}.1 
and give $F_{\overline{H}} \ll 0.9$. 
The relevant parameters of the PYTHIA generator from section
~\ref{sect:interpretation}.2, including the diquark fragmentation 
function, were randomly varied as described 
in the appendix. In these studies, $F_{\overline{H}}$ values less
than 0.86 were not obtained. The experimental result deviates from
this lower limit by 3.8 standard deviations.

The advanced popcorn model has fewer parameters available for tuning
and provides a significantly worse description of data as seen
from the larger $\chi^2$ value.
The larger $\chi^2$ for this model arises to a large extent
from two well known facts. The absolute number of 
$\Lambda \; \overline{\Lambda}$ pairs is too large and the distribution 
of the $\Lambda \; \overline{\Lambda}$ rapidity differences is too narrow  
in comparison to the observation~\cite{lambda-pair-opal}. 
On the other hand, the average $\Lambda \overline{\Lambda}$ 
rapidity difference in the simple popcorn model with a very 
large popcorn parameter is too large, so that a combination of 
the two models might possibly describe the $\Lambda \overline{\Lambda}$
correlation well.

To complete the comparison of correlations with results from Monte 
Carlo generators, the DELPHI 
$\mrm{p} \pi \overline{\mrm{p}}$ ~\cite{ppip-delphi} correlation was 
also investigated. This was done for rapidity ordered 
$\mrm{p} \pi \overline{\mrm{p}}$ and $\pi \mrm{p} \overline{\mrm{p}}$ 
or $\mrm{p} \overline{\mrm{p}} \pi$
particle configurations inside event hemispheres.
The selection cuts and the definition of the minimum rapidity
gap $\Delta y_{min}$ between a selected pion and the next proton 
were taken from \cite{ppip-delphi}.
The discriminating variable of~\cite{ppip-delphi}
is the ratio of intensities
\begin{eqnarray}
R(\Delta y_{min})=\frac{N(\mrm{p} \pi \overline{\mrm{p}})}
{N(\mrm{p} \pi \overline{\mrm{p}})+N(\pi \mrm{p} \overline{\mrm{p}}
+\mrm{p} \overline{\mrm{p}} \pi)}
\end{eqnarray}
at the rapidity difference $\Delta y_{min}$.
A strong dependence on the popcorn effect had been seen
by~\cite{ppip-delphi} at large values of $\Delta y_{min}$. For three
bins in the range $0.625 \le \Delta y_{min} \le 1$, the observed 
distribution agreed with a subsample of Monte Carlo events 
without the popcorn mechanism and disagreed with a disjunct subsample, 
containing popcorn mesons, by more than five standard 
deviations, averaged over the three bins.

These results could be reproduced with the simulations 
described here: The  $R$ distributions obtained with the parameter 
set~ MC1, but without detector corrections, 
agreed with DELPHI's observation within 1.9 standard 
deviations, while there was disagreement between the data and 
Monte Carlo study MC2 by 5.2 standard deviations, averaged over
the same three $\Delta y_{min}$ bins.
Variations of the fragmentation model had not been studied 
in \cite{ppip-delphi}.
An increase of the
fragmentation parameter PARJ(45) to unity, fixing the other parameters
of the simulation MC2, reduces the difference between the data and the 
model prediction to 2.4 standard deviations. 
Furthermore, it was found that the advanced popcorn model with the 
parameter set~MC5 results in almost the same function 
$R(\Delta y_{min})$ as the model MC1 without the popcorn effect. 

The modification of the parameter PAR(45) in the simulation MC2 
increased the $\chi^2$ value in table~\ref{table:MCcorrhyp} from 59 to 70 
due to a shift of $\overline{\Delta y}_{\Lambda \overline{\Lambda}}$
to a value below the observation, without degrading the 
description of the other observables of section~\ref{sect:observables}.
In summary, these results indicate a high sensitivity of the 
rapidity correlations to the fragmentation dynamics.

\begin{table}
\begin{center}
\begin{tabular}{|c|c|r|r|r|r|c|}
\hline
 popcorn parameter   &  &  &  &  &  & \\
 PARJ(5)             &  & \raisebox{1.5ex}[0cm][0cm]{0.} &
    \raisebox{1.5ex}[0cm][0cm]{0.5}  &
    \raisebox{1.5ex}[0cm][0cm]{1.}   &
    \raisebox{1.5ex}[0cm][0cm]{5.}   & \\
\hline
\hline
                     &       &     &     &     &     & MC5 \\
 observable          & data  & MC1 & MC2 & MC3 & MC4 & advanced \\
                     &       &     &     &     &     & popcorn  \\
\hline
protons              & $1.046 \pm 0.026 $ & 1.03  &  1.03 & 1.02 
                     & 1.09 & 1.01 \\
$\Delta^{++}$        & $0.087 \pm 0.033$ & 0.098 &  0.118& 0.122
                     & 0.122 & 0.127 \\
$\Lambda$            & $0.388 \pm 0.009$ & 0.369 &  0.362& 0.366
                     & 0.388 & 0.388 \\
$\Sigma^+ + \Sigma^-$& $0.181 \pm 0.018$ & 0.129 &  0.131& 0.129 
                     & 0.133 & 0.141 \\
$\Sigma^0$           & $0.076 \pm 0.010$ & 0.069 &  0.070& 0.069
                     & 0.070 & 0.073 \\
$\Xi^-$            & 0.$0258 \pm 0.0009$ & 0.029 &  0.029& 0.028
                     & 0.029 & 0.024 \\
$\Sigma^*(1385)^{+,-}$ & $0.046 \pm 0.004$ & 0.043  & 0.048& 0.052
                     & 0.051 & 0.054  \\
$\Omega^-$           & $0.0016 \pm 0.0003$ & 0.0005 & 0.0004& 0.0004
                     & 0.0004 & 0.0004 \\ 
\hline
\end{tabular}
\end{center}
\caption{\it Baryon rates per multihadronic event.
The errors of the simulation are smaller than the last digit shown.}
\label{table:MCobservables}
\end{table}

\section{Conclusions}
\label{sect:conclusion}

Our investigations indicate that the fragmentation models have not yet 
reached a state where they can quantitatively describe the 
$\Lambda\overline{\Lambda}$ correlations, 
the $\mrm{p}\overline{\mrm{p}}\pi$ correlations and the 
$\Sigma^-$ antihyperon correlation $F_{\overline{H}}$ simultaneously.
Neither the $\Lambda\overline{\Lambda}$ nor the 
$\mrm{p}\overline{\mrm{p}}\pi$ correlations can provide a clear conclusion 
about the popcorn effect because they can both be described to an 
acceptable level by models either with and without the popcorn mechanism.
The PYTHIA generator without the popcorn effect can 
reproduce both rapidity correlations simultaneously within two 
standard deviations. 
Both correlations are also in acceptable agreement 
with model predictions including the popcorn effect.
However, we were unable to find a variant of the popcorn model
that could simultaneously describe both types of correlations.

In this work, the mechanism of baryon formation was studied  by 
counting $\Sigma^-$ hyperons and correlated antihyperons 
from hadronic \Zo decays. 
The result $F_{\overline{H}}=0.48 \pm 0.10$, based on 
data taken by the OPAL experiment at LEP, 
favors a large popcorn parameter in the simple popcorn model
and is also consistent with the advanced popcorn model.
The fragmentation dynamics play no role here
because the rapidity is not used in the analysis.
Correlated particle momenta play an indirect role 
only, because they influence the detection efficiencies,
but the final result contains a correction and a systematic 
error for mismodeling.

Trivial correlations between hyperons and antihyperons
based on baryon number and strangeness conservation,
as predicted by the thermodynamic model, always exist.
Due to  the limited data statistics and the insensitivity to the 
dynamics, our result gives only a weak indication for 
non-trivial quark correlations between baryon-antibaryon pairs. 
Statistical models, only constrained by conservation laws,
differ at most by 2.6~standard deviations.

The result for $F_{\overline{H}}$ deviates from the 
lower limit of simulations without the popcorn effect by 
3.8~standard deviations and thus demonstrates the need for the
the popcorn effect in order to reproduce baryon correlations
within the diquark fragmentation model, where a baryon and an 
antibaryon share two valence quark-antiquark pairs. More generally, 
one expects any fragmentation model with very strong valence quark 
correlations between baryons and antibaryons to be disfavored.

\section*{Appendix: Tuning of the Monte Carlo generator}
\begin{table}
\begin{center}
\begin{tabular}{|c|c|c|c|c|c|c|}
\hline
     & standard      &      &        &      &      & MC5  \\
\raisebox{1.5ex}{model}     & OPAL   & MC1  
                            &  MC2   & MC3  & MC4  & advanced \\
\raisebox{1.5ex}{parameter} & tune   &      
                            &        &      &      & popcorn  \\        
\hline
\hline
PARJ(5)    & 0.5   & 0.    & 0.5   & 1.    & 5.    & - \\
\hline
PARJ(1)    & 0.085 & 0.092 & 0.100 & 0.101 & 0.117 & 0.226 \\
PARJ(2)    & 0.31  & 0.33  & 0.33  & 0.34  & 0.34  & 0.31 \\
PARJ(3)    & 0.45  & 0.32  & 0.41  & 0.40  & 0.48  & - \\
PARJ(4)    & 0.025 & 0.055 & 0.029 & 0.021 & 0.007 & - \\
PARJ(6)    & 0.5   & -     & 0.33  & 0.56  & 0.35  & - \\
PARJ(7)    & 0.5   & -     & 0.21  & 0.20  & 0.28  & - \\
PARJ(8)    & -     & -     &  -    & -     & -     & 1.00 \\
PARJ(9)    & -     & -     &  -    & -     & -     & 2.07 \\
PARJ(18)   & -     & -     &  -    & -     & -     & 0.18 \\
PARJ(45)   & 0.5   & 0.16  & 0.22  & 0.52  & 0.33  & 0.27 \\
\hline
\end{tabular}
\end{center}
\caption{\it Parameter sets used for the simulations in tables 
\ref{table:MCcorrhyp} and \ref{table:MCobservables}.}
\label{table:MCparameter}
\end{table}

\begin{table}
\begin{center}
\begin{tabular}{|r|r|r|r|r|r|}
\hline
 MC parameter & PARJ(1)& PARJ(2)   & PARJ(3)   & PARJ(4)   & PARJ(45) \\
\hline
\hline
mean value &      0.091&      0.31 &      0.35 &      0.045&       0.50 \\
interval   &$\pm$ 0.008&$\pm$ 0.04 &$\pm$ 0.08 &$\pm$ 0.012& $\pm$ 0.50 \\
\hline
\end{tabular}
\end{center}
\caption{\it Parameter ranges used for random generation of
Monte Carlo parameter sets without the popcorn effect.}
\label{table:parscan}
\end{table}

The tuning started with the PYTHIA parameters of~\cite{opaltunejt74},
listed as ``standard OPAL tune'' in table~\ref{table:MCparameter}.
In a first step, the parameters, except the popcorn parameter, were varied
individually and minimum values of $\chi^2$ were searched for.
This process was iterated. Search ranges for all parameters
were defined either by requiring a maximum increase in $\chi^2$ of 20
or, in the case of smaller changes, by allowing a parameter shift 
of $\pm 100$\%.
Finally, all parameters were varied randomly within these ranges
to search for $\chi^2$ values lower than that of the solution already found. 
Between 200 and 300 random parameter sets were generated at fixed PARJ(5) 
for the final search and $10^5$ events were generated per parameter set. 
The results for the selected parameters sets MC1 to MC5 are based on 
$10^6$ events. The contributions of the four classes of observables to 
$\chi^2$ at the minimum value 56 are about 40 from the baryon rates, 
10 from the meson rates and 6 from the remaining six observables.
 
The measured and simulated baryon rates per \Zo~decay are given in
table~\ref{table:MCobservables}, while the corresponding PYTHIA 
parameters are listed in table~\ref{table:MCparameter}. 
The $\Omega^-$ baryon was included in the optimization to avoid its 
almost complete suppression. The results for the 
parameter sets MC3 and MC5 are similar to those in our previous
publication~\cite{lambda-pair-opal}, where the case without the popcorn
effect was not investigated.
The reproduction of the meson sector does not change much during 
the tuning and is therefore omitted from table~\ref{table:MCobservables}.

Finally, table ~\ref{table:parscan} gives the parameter space
for the study of the parameter dependence of $F_{\overline{H}}$.
All cases with $\chi^2<110$ were kept, allowing average
discrepancies between the data and the model as large as those
obtained with the modified popcorn model.

\section*{Acknowledgments}
We are grateful to F.Becattini for applying
the thermodynamic model to the $\Sigma^-$ antihyperon correlations
and we thank J.Lamsa and K.Hamacher from the DELPHI 
collaboration for discussions about the proton-antiproton-pion 
correlations.
We particularly wish to thank the SL Division for the efficient 
operation of the LEP accelerator at all energies
 and for their close cooperation with
our experimental group.  In addition to the support staff at our own
institutions we are pleased to acknowledge the  \\
Department of Energy, USA, \\
National Science Foundation, USA, \\
Particle Physics and Astronomy Research Council, UK, \\
Natural Sciences and Engineering Research Council, Canada, \\
Israel Science Foundation, administered by the Israel
Academy of Science and Humanities, \\
Benoziyo Center for High Energy Physics,\\
Japanese Ministry of Education, Culture, Sports, Science and
Technology (MEXT) and a grant under the MEXT International
Science Research Program,\\
Japanese Society for the Promotion of Science (JSPS),\\
German Israeli Bi-national Science Foundation (GIF), \\
Bundesministerium f\"ur Bildung und Forschung, Germany, \\
National Research Council of Canada, \\
Hungarian Foundation for Scientific Research, OTKA T-038240, 
and T-042864,\\
The NWO/NATO Fund for Scientific Research, the Netherlands.\\


\end{document}